\documentclass[twocolumn,secnumarabic,amssymb, nobibnotes, aps, prd, superscriptaddress, longbibliography]{revtex4-1}

\makeatletter
\def\@bibdataout@aps{%
\immediate\write\@bibdataout{%
@CONTROL{%
apsrev41Control%
\longbibliography@sw{%
    ,author="48",editor="1",pages="1",title="0",year="1"%
    }{%
    ,author="48",editor="1",pages="1",title="",year="1"%
    }%
  }%
}%
\if@filesw \immediate \write \@auxout {\string \citation {apsrev41Control}}\fi 
}
\makeatother

\usepackage{siunitx}
\usepackage{booktabs}
\usepackage{graphicx}
\usepackage{makecell}
\usepackage{amsmath}
\usepackage{float}
\usepackage{multirow}
\begin{document}

\title{Electronic structure of noncentrosymmetric B20 compound HfSn and tuning of multifold band-crossing points}%

\author{Dijana Milosavljevi\'{c}}%
\email[]{dijana.milosavljevic@mpi-halle.mpg.de}
\affiliation{Max Planck Institute of Microstructure Physics, 06120 Halle (Saale), Germany}
\author{Helge Rosner}
\affiliation{Max Planck Institute for Chemical Physics of Solids, 01187 Dresden, Germany}
\author{Annika Johansson}
\affiliation{Max Planck Institute of Microstructure Physics, 06120 Halle (Saale), Germany}
\begin{abstract}
We present a detailed theoretical study of the electronic structure of hafnium tin HfSn crystallizing in a B$20$ structure, renowned for the diversity of physical and peculiar topological properties. The chiral crystal structure of these materials protects multifold band crossings located at high symmetry points. We employ density functional methods to reveal basic features of the band structure and Fermi surface topology of HfSn, on top of which an effective tight-binding model is built. The compound exhibits a fourfold band crossing pinned at the $\Gamma$ point. We investigate routes that can shift such crossings towards the Fermi level, offering a unique way to possibly tune the compound's properties. Specifically, we show that the energy position of the fourfold crossing can be easily manipulated via external perturbations such as strain and pressure. Considering that this point carries a topological charge larger than 1, such tuning is of great importance. We anticipate that the approach presented in the current study can be utilized to investigate symmetry protected crossings in a wide class of materials.
\end{abstract}
\maketitle

\section{Introduction}
\hspace*{2em} Synthesis and investigation of functional materials not available in nature are imperative for new technological developments and advancement in various disciplines. An important criterion is that the crystal structures of these novel materials are as simple as possible but still host a wide spectrum of peculiar physical and chemical properties. This condition is fulfilled in the compounds with the noncentrosymmetric B$20$ crystal structure, which has a simple cubic unit cell and as symmetry elements screw and 3-fold rotations only \cite{fecher}. The B$20$ family, named after their crystal structure, is an exciting class of materials in condensed matter physics. Among the representatives of this family, there are metals \cite{CrSimetal}, semiconductors \cite{kondoinsulator}, Kondo insulators \cite{kondoinsulator} and superconductors \cite{aubesuperconductor}. Besides the conventional paramagnetic, diamagnetic, and ferromagnetic state, the B$20$-type members can exhibit chiral magnetic order featuring nontrivial chiral vortex-line spin textures and skyrmions \cite{mnsiskyrmions}. Moreover, 'nonmagnetic' ones display three, four, and sixfold degenerate band crossings pinned at high symmetry points with topological charges greater than unity \cite{maximalchernnumber}.  On top of the intriguing physical properties, few members have already been recognized as stable electrocatalysts for hydrogen evolution reaction \cite{qunyang}. \\
\hspace*{2em} Whereas most of the literature on the B$20$ compounds focuses on monosilicides and monogermanides with transition metals from the seventh and eighth groups of the periodic table \cite{severin}, scant attention has been devoted to the B$20$ structure compound with a transition metal of the fourth group. The first known compound associated with this particular subclass of the rich B$20$ family is HfSn \cite{schob}. In the light of extraordinary mechanical, reflectory, and high-corrosion resistance properties of Hf, investigation of its binary compounds is valuable. It is noteworthy that HfC is in the list of the most refractory materials under atmospheric pressure predicted to have the highest melting points \cite{highestmeltingpoint}. In addition, the monolayers of HfC are predicted to be quantum spin Hall insulators by applying external in-plane strain, qualifying them as potential candidates for spintronics applications \cite{strain}. \\
\hspace*{2em} The wealth of existing applications of hafnium and its binary compounds and the uniqueness of the B20 family motivated us to investigate systematically the properties of HfSn. To the best of our knowledge, there are no experimental reports on the electronic structure properties of this compound. While in the literature there exist density functional theory (DFT) calculations of the band structure of HfSn \cite{opticalpaper}, many features associated with it have not been established yet, opening a space for further investigation. Also, very little attention has been dedicated to the other two four-group transition metal B$20$ representatives, HfSb and ZrSb. As an illustration, HfSb has been proposed to be a semimetal exhibiting the quantum anomalous Hall phase \cite{huangnanoscale}, and  ZrSb is an insulator found to be trivial \cite{huangnanoscale}.\\
\hspace*{2em}  In the present work, we report an extensive computational study of HfSn by means of DFT and subsequent modelling approach. The so-derived effective Hamiltonian is employed for the investigation of the topological properties of this material. An ultimate goal of the paper is to construct a clear physical picture of the electronic structure of HfSn, contributing to a better comprehension of B$20$ compounds, in particular, those with a transition metal from the fourth group, under-examined so far.\\
\hspace*{2em} The HfSn phase was first reported by Schob \emph{et al.} \cite{schob} to be stable over a narrow temperature range around $1000^{\circ}$C. In the subsequent publication of Tsyganova \emph{et al.} \cite{tsyganova}, the Hf-Sn phase diagram, constructed based on the differential thermal analysis and X-ray diffraction results, did not contain the HfSn phase. Contrasted with this result, Romaka \emph{et al.} \cite{romaka} established the binary Hf-Sn system phase diagram containing HfSn, Hf$_{5}$Sn$_{4}$, HfSn$_{2}$ and Hf$_{5}$Sn$_{3}$. Further experiments on related Hf-Ni-Sn systems validated the existence of the HfSn phase (Stadnyk \emph{et al.} \cite{stadnyak}). Tang \emph{et al.} \cite{tang} disregard the HfSn phase in the thermodynamic modelling but still report the first principle results of its crystal structure, indirectly pointing to the occurrence of HfSn. It is important to emphasize that in the current paper, we do not address the stability of the HfSn phase. Instead, we take the first crystallographic report by Schob \emph{et al.} \cite{schob} and employ it as a starting point for all calculations presented in the current work.\\
\hspace*{2em} The work is organized as follows. In Section \ref{methods} we establish the theoretical methods utilized in the current study. Section \ref{crystalstrcuture} contains a brief description of the experimental crystal structure \cite{schob}. In \ref{resultsection} \ref{electronicstructurecalculations}, we report the electronic structure of HfSn obtained by means of DFT. We use this information to explore how external perturbations like hydrostatic pressure (Section \ref{resultsection} \ref{pressure}), uniaxial strain and doping (Section \ref{resultsection} \ref{Uniaxialstrainandchemical pressure}) can be utilized to tune points of interest in the electronic structure, namely multifold band crossings. In Section \ref{tightbindingmodel} we construct an effective tight binding model used then to investigate the topological properties of these points.

\section{Methods}
\label{methods}
\hspace*{2em} The calculations are performed by employing the local density approximation (LDA) and generalized gradient approximation (GGA) for the exchange-correlation functional as implemented in the full potential local orbital (FPLO) code \cite{fplo} and Vienna Ab Initio Simulation Package  (VASP) \cite{vaspcitaton1} \cite{vaspcitation2} \cite{vaspcitation3}, within the projector-augmented-plane-wave (PAW) method. We use Perdew-Wang \cite{perdewwang} and Perdew-Burke-Ernzerhof \cite{perdewburke} for LDA and GGA, respectively. For VASP, we use parametrization of Ceperly Alder \cite{ceperlyadler} for the LDA, while GGA parametrization is the same as in the FPLO code.  The spin-orbit coupling (SOC) is treated nonperturbatively in FPLO and in a second variational scheme in VASP. Starting from the experimental crystal structure \cite{schob}, we relax the atomic positions until the forces reach value less than $10^{-3}$ eV/$\si{\angstrom}$. In addition to the atomic positions, we fully relax the volume of the unit cell of the HfSn. We downfold our FPLO calculations on a Wannier functions (WFs) basis incorporating Hf $5$d, $6$p, and Sn $5$p orbitals, with the requirement of maximum localizations. Futhermore, we use the output of our bulk VASP calculations to construct the maximally localized (ML) WFs employing the Wannier90 \cite{wannier90citation1} code. The so obtained MLWFs are used to calculate the total and band resolved Berry curvature. The topological charges are calculated utilizing the WannierTools package \cite{wanniertools}. We plot the band resolved Berry curvature with the help of WannierBerri code \cite{wannierberri}. A well-converged $k$-mesh of 12x12x12 $k$-points in the whole Brillouin zone is employed. For VASP calculations, we set the $k$-point grid to 8x8x8, and the convergence of the total energy is $10^{-6}$ eV.
\section{Crystal structure}
\label{crystalstrcuture}
\begin{figure}[h!]
\includegraphics[width=0.5\textwidth, keepaspectratio]{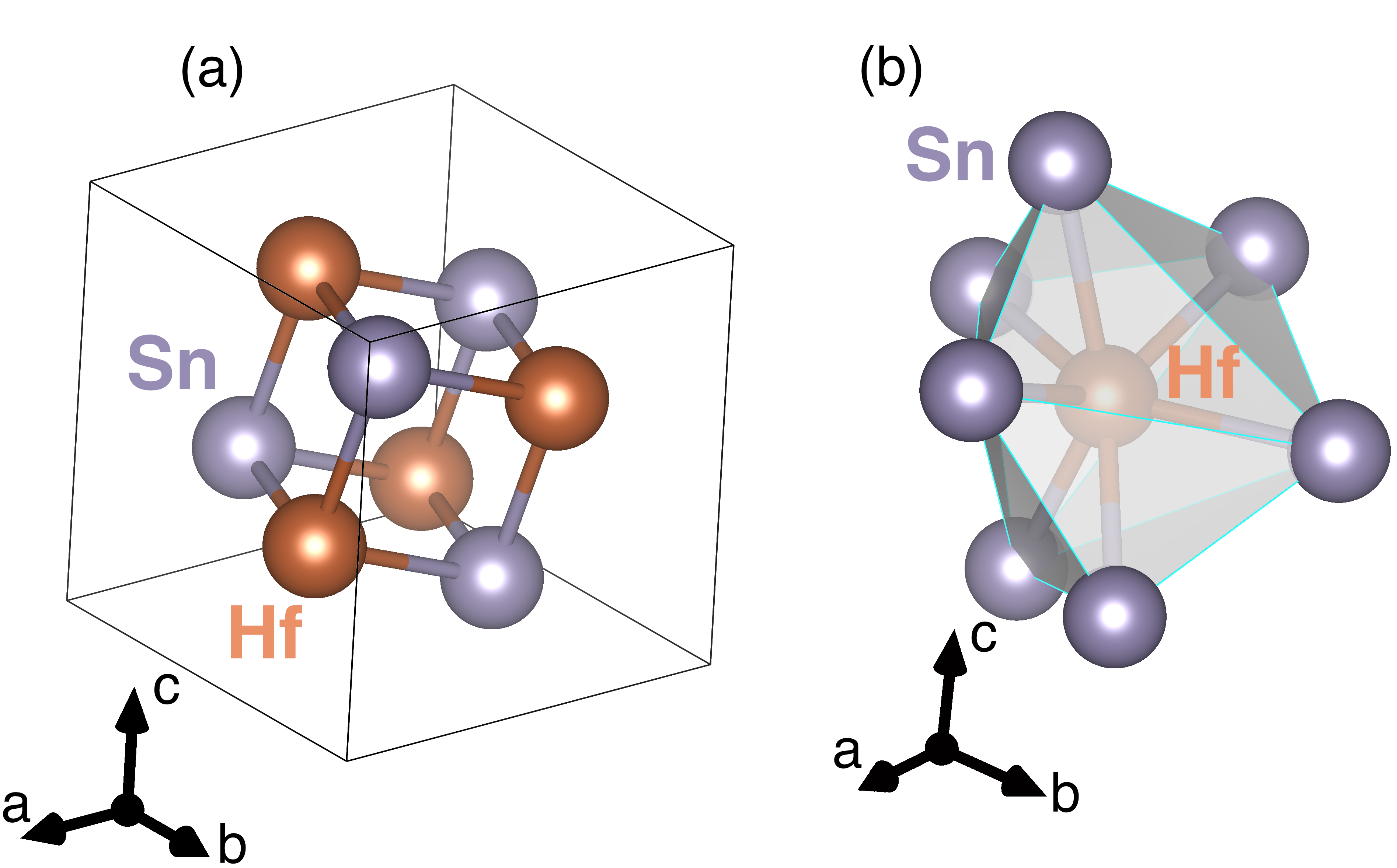}
\caption{(a) The unit cell of the B20 crystal structure of HfSn. The orange atoms in the unit cell depict Hf atoms and purple atoms illustrate Sn atoms. (b) Coordination polyhedron of an Hf atom. Six Sn atoms are located at the distance of $2.99$  $\si{\angstrom}$ and one is at $3$ $\si{\angstrom}$ (or $7$ equidistant, depending on error bars).}
\label{figure1}
\end{figure}

\hspace*{2em} The crystal structure of HfSn is described by a B$20$ Strukturbericht designation, a name that accounts for binary compounds that crystallize in the space group P2$_{1}$3 (number 198 in the International Table). A distinguished feature of this family is the lack of mirror and inversion symmetry, resulting in a chiral crystal structure. Thus, in total, this space group contains $12$ symmetry operations: $2$-fold and $3$-fold screw rotations and simple $3$-fold rotations.  In a simple cubic unit cell, there are four atoms of each species located at the 4$a$-type Wyckoff positions determined by the parameters x$_{\text{Hf}}$ $=$ $0.155$ and x$_{\text{Sn}}$ $=$ $0.845$ \cite{schob} for Hf and Sn atoms, respectively (see Fig. \ref{figure1}). The experimental room temperature lattice parameter is $a=5.594$ $\si{\angstrom}$ \cite{schob}. Each atom, Hf and Sn, is bonded to 6 nearest neighbours (NNs) of opposite kinds at a distance of $2.99$ $\si{\angstrom}$ and one at $3.00$ $\si{\angstrom}$. If one takes into account experimental uncertainty, there are seven equidistant NNs at a distance $a\sqrt(3)/(1+\sqrt(5))$$=$$2.99$ $\si{\angstrom}$ resulting in an 'ideal' B$20$ structure \cite{vocaldo}. The next nearest neighbours (NNNs) of each atomic species are atoms of the same kind whose separation is equal to $3.45$ $\si{\angstrom}$ (for all six atoms). 

\section{Results}
\label{resultsection}
\subsection{Electronic structure}
\label{electronicstructurecalculations}
\begin{figure}[h!]
\includegraphics[width=0.5\textwidth, keepaspectratio]{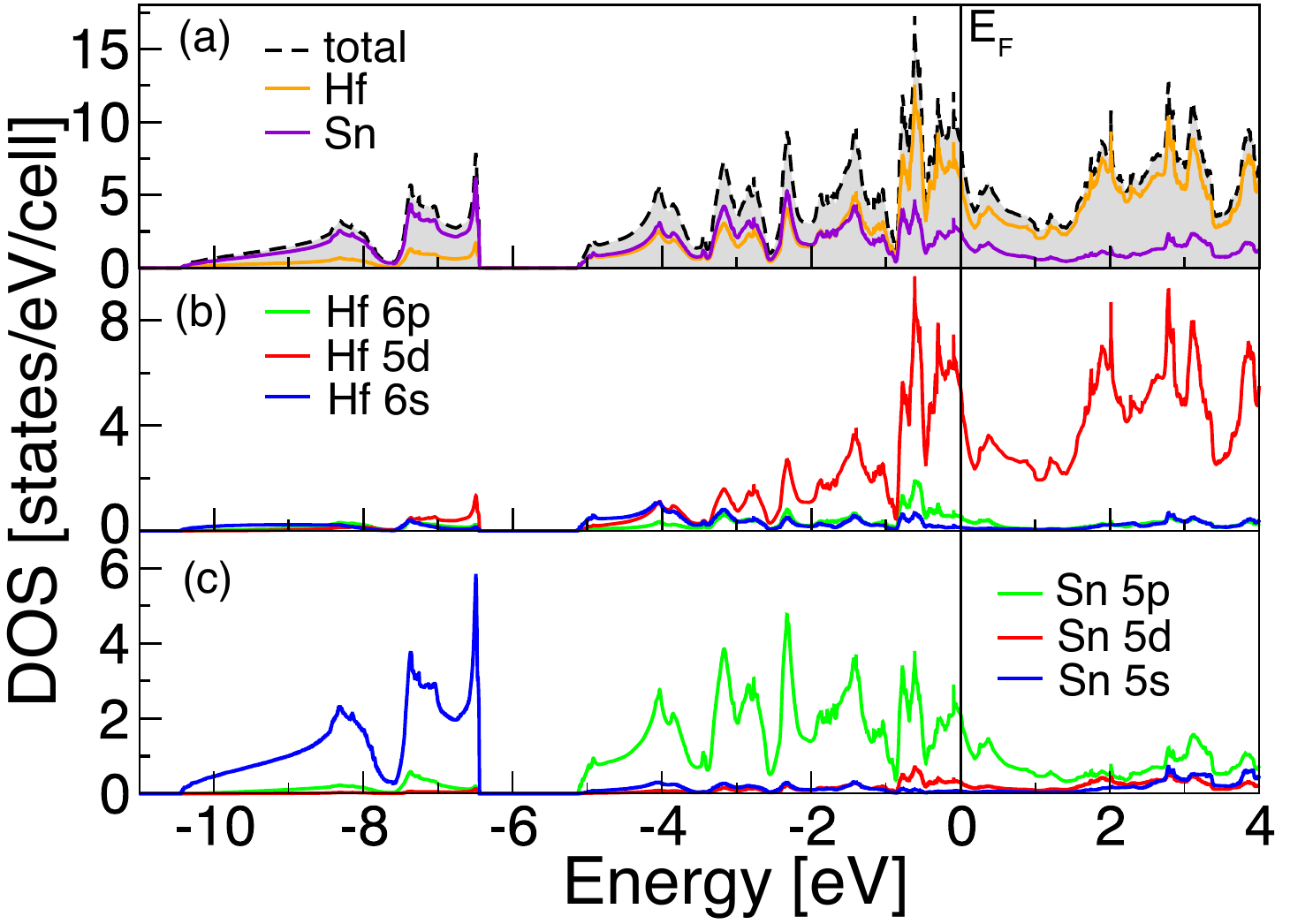}
\caption{Calculated LDA total and atom resolved partial electronic densities of states of the HfSn ((a), (b), (c)). The Fermi level is at zero energy and marked by a full black line.}
\label{figure2}
\end{figure}

\hspace*{2em}  In the first place, we present the analysis of the density of states (DOS) and electronic band structure. In Fig. \ref{figure2} the total and atom resolved density of states of HfSn are shown. From the top panel (a), it can be deduced that the density of states at the Fermi level E$_{F}$ is approximately $9.0$ states/(eV $\cdot$ cell). This result is comparable to $10.5$ states/(eV $\cdot$ cell) obtained for CrGe \cite{ulrichcrg} with the FPLO code. In both systems, such a high value of DOS is caused by the presence of many bands crossing the Fermi level. The valence band complex is dominated by Sn states with sizeable contribution of Hf states above $-1.5$ eV (see purple and orange lines in panel (a) of Fig. \ref{figure2}. 
\begin{figure}[h!]
\includegraphics[width=0.52\textwidth, keepaspectratio]{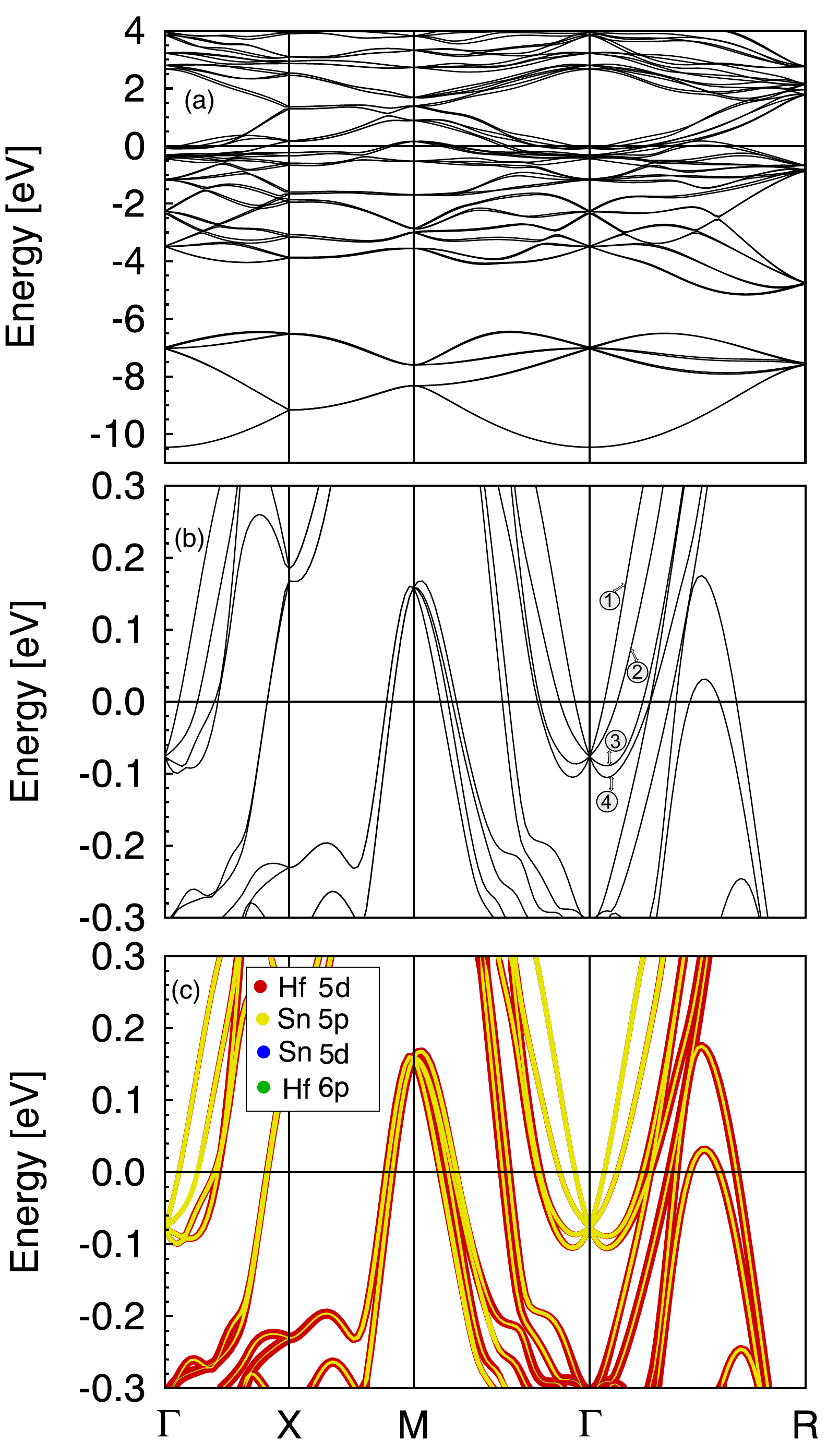}
\caption{(a) Calculated LDA band structure of HfSn with SOC; (b) Zoom of the band structure around the Fermi level E$_{F}$; Numbers 1,2,3,4 denote the bands forming fourfold degenerate point just below the E$_{F}$ (c) Calculated orbital band character in the vicinity of E$_{F}$: Hf $5$d (red), Hf $6$p (green), Sn $5$d (blue) and Sn $5$p (yellow).  }
\label{bulkbandstructure}
\end{figure}
\begin{table*}[htbp]
\centering
\begin{ruledtabular}
\begin{tabular}{lcccccccccc}
\toprule
&\multicolumn{3}{c}{VASP} & \multicolumn{2}{c}{FPLO} & \multicolumn{2}{c}{exp \cite{schob} VASP} & \multicolumn{2}{c}{exp \cite{schob} FPLO} \\
\cmidrule(r){1-3} \cmidrule(r){4-5}\cmidrule(r){6-7}\cmidrule(r){8-9}
& LDA   &    GGA      &      &  LDA   &  GGA & LDA  & GGA  & LDA & GGA\\
\midrule
a &   5.5368  &   5.6544     &        &  5.5555  & 5.6817& \multicolumn{2}{c}{5.594} &  \multicolumn{2}{c}{5.594} & \\
d$_{1}$ & 2.882 x1     &   2.935 x1         &          &  2.890 x1 &  2.949 x1& \multicolumn{2}{c}{\multirow{3}{*}{2.990}}  & \multicolumn{2}{c}{\multirow{3}{*}{2.990}}   &\\
d$_{2}$  &  2.917 x3  &     2.979 x3        &           &   2.926 x3 & 2.992 x3& & & & \\
d$_{3}$ &   3.045 x3   &    3.115  x3        &          &   3.058 x3 & 3.131 x3 & & & &\\
t            &   -41.17    &   -104.13         &          &  -56.12  & -135.65 & -51.10 & -72.41& -78.33 &-102.27\\
V &  169.737    &      180.784      &          &171.463    & 183.415 & \multicolumn{2}{c}{175.052}  &  \multicolumn{2}{c}{175.052} &\\
x$_{\text{Hf}}$& 0.144   & 0.143     &      & 0.144      & 0.143    & \multicolumn{2}{c}{0.155} & \multicolumn{2}{c}{0.155} &\\
x$_{\text{Sn}}$& 0.843   & 0.844     &      & 0.843      & 0.843    &\multicolumn{2}{c}{0.845} & \multicolumn{2}{c}{0.845}  &\\
\bottomrule     
\end{tabular}  
\label{distancesandlatticeparameters}
\end{ruledtabular}
\caption{Lattice parameter a [$\si{\angstrom}$], distances d$_{1}$, d$_{2}$ and d$_{3}$ [$\si{\angstrom}$], energy separation t [meV] and volume V [$\si{\angstrom}^{3}$] obtained by full structural optimization employing the FPLO and VASP codes with various exchange-correlation functionals (LDA versus GGA). d$_{1}$, d$_{2}$ and d$_{3}$ denote the NN distances between Hf and Sn atoms (d$_{1}$ appearing once and d$_{2}$ and d$_{3}$ three times), "t" stands for the energy separation of the degenerate points at $\Gamma$ from the Fermi level, and V is the volume obtained after structural optimization. The Wyckoff positions for Hf and Sn atoms are denoted with x$_{\text{Hf}}$ and x$_{\text{Sn}}$, respectively. The values for the experimental crystal structure (exp \cite{schob}) are also presented.}
\label{tablewithlp}
\end{table*}  
The orbital resolved DOS shown in the panels (b) and (c)  of Fig. \ref{figure2} indicate that Hf and Sn $5$d states contribute mainly in the energy window from $-5$ to $4$ eV. The same also applies to $6$p and $5$p block of Hf and Sn, respectively. Specifically, the Hf $5$d states contribute about two-thirds of states at the Fermi level E$_{F}$ and Sn $5$p about one-quarter. Additionally, the Hf $6$p states are present at E$_{F}$ with $6$ $\%$ compared to the total DOS and Sn $5$d states with 3 $\%$. The contribution of all other states is about $1$ $\%$ and therefore negligible. This analysis is of major importance because it guides the construction of an effective tight-binding model. \\
\hspace*{2em}We stress that two different exchange-correlation functionals (LDA versus GGA) qualitatively lead to the same results. Also, two codes (FPLO and VASP) yield the same qualitative behaviour. Hence, in the discussion below, we focus mainly on the FPLO results. The bulk band structure of HfSn for the experimental crystal structure \cite{schob} is shown in Fig. \ref{bulkbandstructure}. The overall shape and dispersion of the bands resemble that previously calculated for HfSn \cite{opticalpaper}, providing a justification for our calculations. However, in addition to this, we dedicate special attention to the peculiarities of the band structure like band crossings and their energy separation from the Fermi level. This is an important issue to revisit because it could possibly offer a mechanism to modify the compound properties by external influence. In the energy range from -5.3 to 1.4 eV, in overall there are 30 bands (easiest to see along the $\Gamma$M direction). These states group into sets of subbands which become two and fourfold degenerate in the Brillouin zone center ($\Gamma$ point). There exist five subsets of twofold and fourfold degeneracies. At the Brillouin zone corner (R point), the bands exhibit peculiar two and sixfold degeneracies. At the Brillouin zone boundary, all states are twofold degenerate. The presence of multifold degenerate bands at high symmetry points is a distinguished characteristic of the whole family of B$20$ compounds. This feature is a direct consequence of the structural chirality of the respective real space lattice and the peculiar symmetries of the B$20$ compounds. Owing to the nonsymmorphic $C_{2}$ rotation and time reversal symmetry, the analysis of the linearized \textbf{k}$\cdot$\textbf{p} Hamiltonian reveals the stability of the degeneracies up to higher orders in \textbf{k}$\cdot$\textbf{p} \cite{bradlyn}. The multiplicity of band crossings on high-symmetry points is determined by the dimensionality of the irreducible representation of the little groups \cite{nico}. For the case with SOC, at $\Gamma$ there are representations of dimensionality 2 and 4, and at R, 2 and 6, resulting in twofold and fourfold degeneracies at $\Gamma$, and twofold and sixfold degeneracies at R, in agreement with our calculations.    \\
 \hspace*{2em} A zoom around the Fermi level of the band structure at the $\Gamma$ point reveals the fourfold degeneracy located $78$ and $102$ meV below the Fermi level in LDA and GGA, respectively (see Fig. \ref{bulkbandstructure} (b) for LDA and Appendix \ref{LDAversusGGAcomparison} for GGA, Fig. $\ref{LDAGGAwithandwithoutSOC}$). These degenerate points are more than a factor of two closer to the Fermi level compared to the calculations without SOC (see the Appendix \ref{LDAversusGGAcomparison}, Fig. $\ref{LDAGGAwithandwithoutSOC}$). At the R point, the sixfold degenerate point is significantly lower in energy in both LDA and GGA than the degenerate point at the $\Gamma$. Hence, we mainly focus on the properties of the electronic structure around the $\Gamma$ point in the following. We find the energy position of the degenerate point at $\Gamma$ to be sensitive on the optimization of the experimental crystal structure \cite{schob} (see below). The states in the energy region from $-0.3$ to $0.3$ eV are characterized by a strong hybridization between Hf $5$d and Sn $5$p states (Fig. \ref{bulkbandstructure} (c)). The d-p admixture is frequently encountered in transition metal intermetallic compounds, governing their electronic properties. \\
\begin{figure*}[htbp]
\includegraphics[width=\textwidth, keepaspectratio]{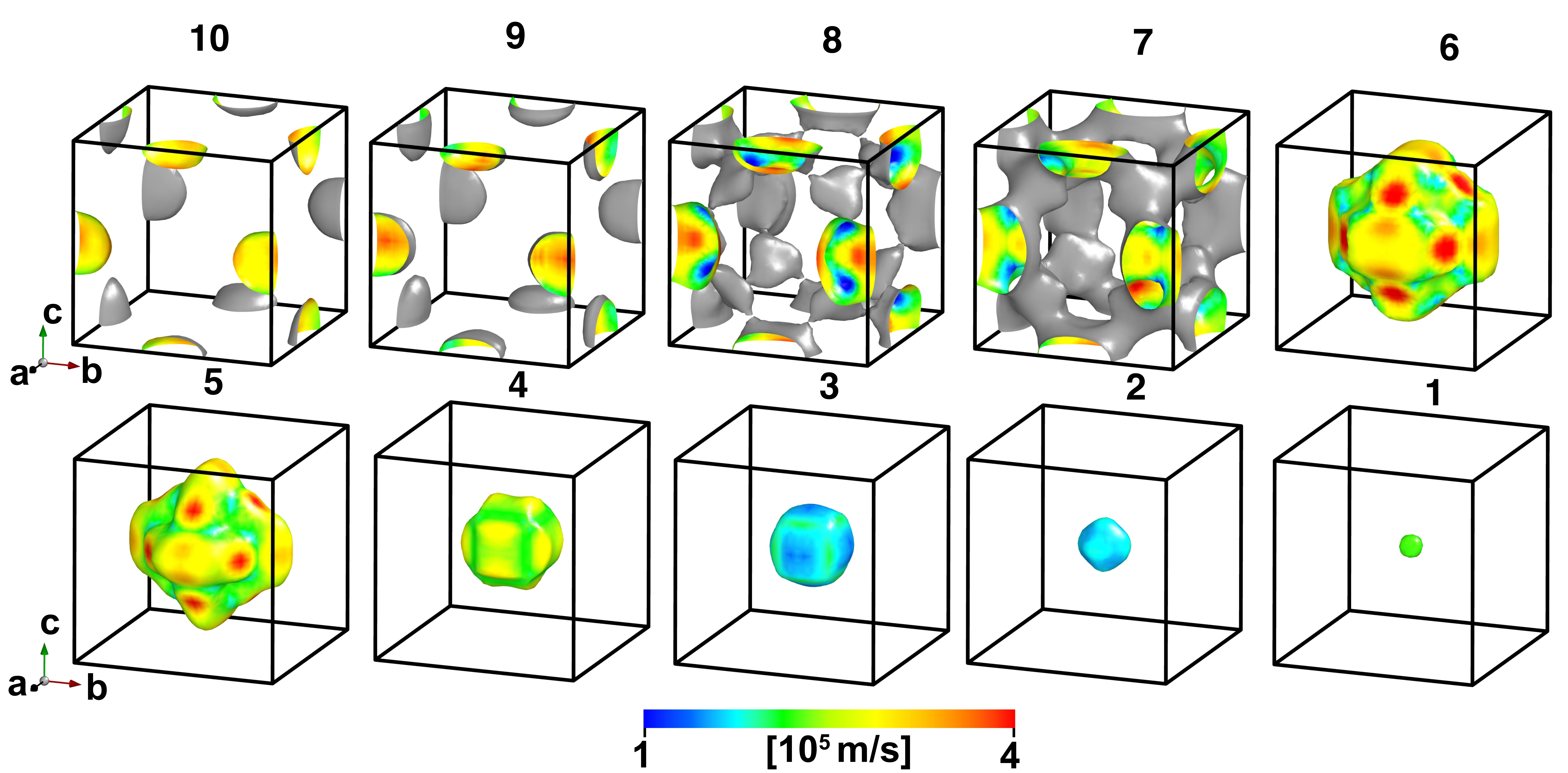}
\caption{The Fermi surfaces of HfSn with the Fermi velocities at various pockets depicted with color gradients. The calculations are presented for the experimental crystal structure with SOC. }
\label{fermisurfacesforhfsn}
\end{figure*}
\hspace*{2em} Our calculations for HfSn, with SOC included, reveal 10 Fermi surface sheets as depicted in Fig. $\ref{fermisurfacesforhfsn}$ (labeled with numbers from $1$ to $10$). Six out of ten Fermi sheets are electron pockets centered at $\Gamma$ (two of them (labelled as $6$ and $5$ in Fig. $\ref{fermisurfacesforhfsn}$) are of the octahedral shape, and four ($4$, $3$, $2$, $1$ in Fig. $\ref{fermisurfacesforhfsn}$) are approximately spherical). Spherical sheets have moderately lower Fermi velocities compared to other sheets. In addition, there is one sheet that forms the webbed tunnel ($7$ in Fig. $\ref{fermisurfacesforhfsn}$), with the remaining three ($10$, $9$, $8$) representing hole pockets located at the edges of the Brillouin zone.  \\
\hspace*{2em} As a next step, we perform structural optimization of the experimental structural data \cite{schob}. The optimization of the atomic coordinates only is followed by a full optimization of the unit cell. These calculations lead to an optimized structure which differs from the "ideal B$20$" structure in the sense that the seven nearest neighbour distances (see Fig. $\ref{figure1}$) are not equal, even not approximately. Instead there are three distinct distances denoted d$_{1}$, d$_{2}$ and d$_{3}$, respectively (see Table \ref{tablewithlp}). We find that the optimization of the crystal structure influences predominately the position of the degenerate point, irrespectively of the basis set employed (plane wave as implemented in VASP versus local orbitals in FPLO). Full optimization of the crystal structure moves the multifold point at $\Gamma$ towards the Fermi level by approximately $22$ meV in LDA (FPLO). In GGA (both codes), this behaviour is opposite and the point moves away from the Fermi level by almost 33 meV compared to the experimental crystal structure \cite{schob} (see Table \ref{tablewithlp}). The difference in the optimized lattice parameters of about $0.02$ $\si{\angstrom}$ in LDA and $0.03$ $\si{\angstrom}$ in GGA between FPLO and VASP can be ascribed thoroughly to the different types of basis sets. Optimization of the atomic coordinates solely (not shown in Table \ref{tablewithlp}) also has an impact on the position of the degenerate point, just with a less pronounced shift, reduced approximately by a factor of two compared to the case when the structure is fully optimized. For further analysis, we make use of our LDA optimized parameters, which can be justified by the fact that LDA performs the best for elemental Sn over the GGA functionals \cite{wucu}. Furthermore, it is established that the tendency of LDA to underestimate the lattice constants is smaller than the GGA overestimation of parameters \cite{ldaunderestimation}. \\
\subsection{Hydrostatic pressure}
\label{pressure}
\begin{figure}[h!]
\includegraphics[width=0.48\textwidth, keepaspectratio]{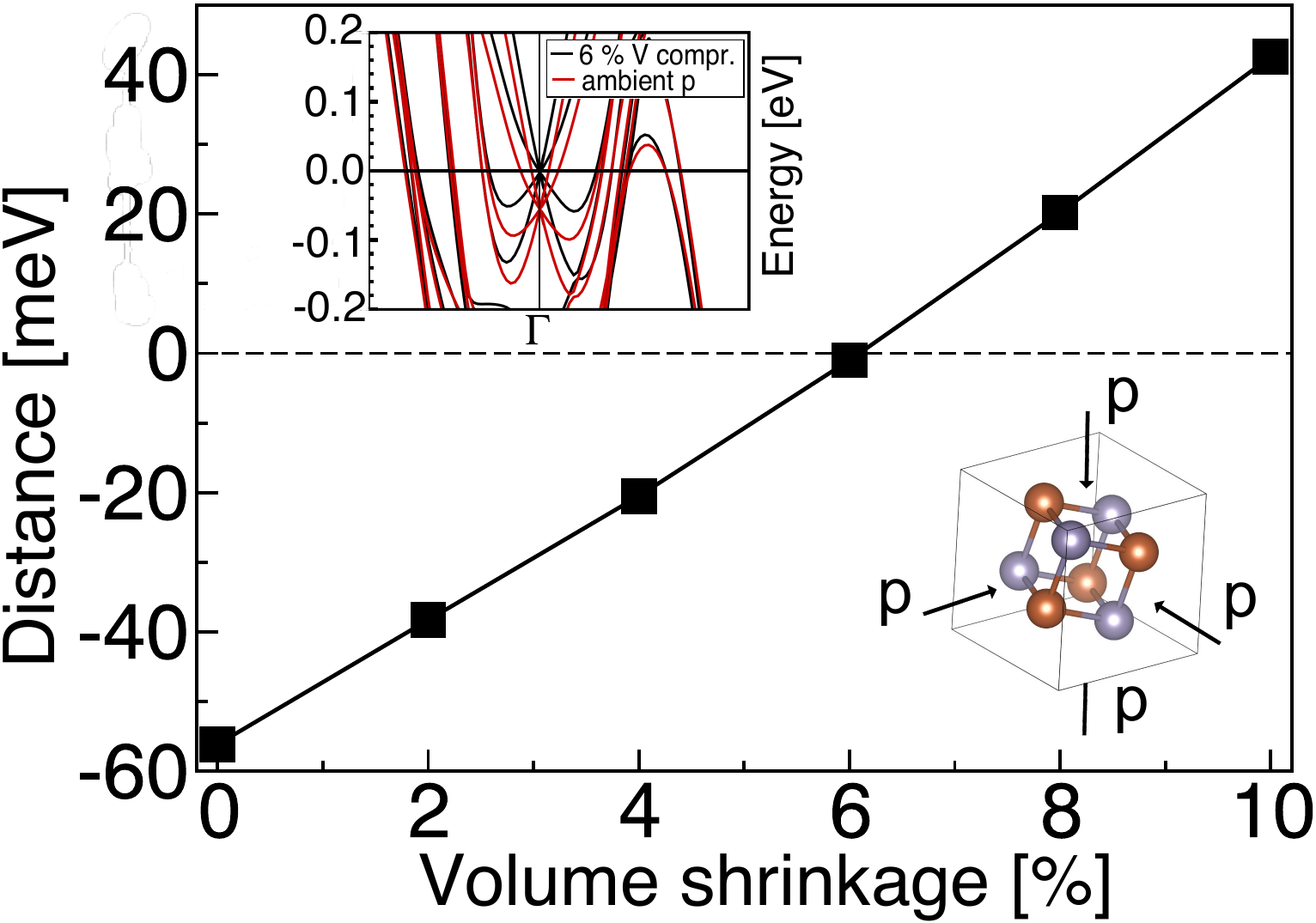}
\caption{Distance of the degenerate point at $\Gamma$ from the Fermi level as a function of the volume compression by hydrostatic pressure application. Inset: Band structure at ambient pressure and $6\%$ of volume reduction coresponding to hydrostatic pressure of about 8.5 GPa.}
\label{hydrostaticpressurefigure}
\end{figure}

\hspace*{2em} From Table \ref{tablewithlp}, it can be concluded that smaller lattice parameters in LDA and consequently smaller volumes lead to a shift of the degenerate point closer to the Fermi level. In a material it is desirable that such band degeneracies occur in the vicinity or exactly at the Fermi level, in order to get a chance to investigate peculiar transport and topological properties originating from the electronic states around these points. Hence, it is of the uppermost importance to identify all possible routes by which such band crossing can be located at E$_{F}$.  Our results illustrate that volume compression acts in favour of achieving this effect. In reality, volume reduction can be achieved by application of hydrostatic pressure. Therefore, this result prompts an investigation of the hydrostatic pressure influence on the properties of HfSn, in particular the position of the fourfold degenerate point at $\Gamma$.  In the following, we report on the hydrostatic compression studies on HfSn. In Fig. \ref{hydrostaticpressurefigure}, we show the separation of the degenerate point at $\Gamma$ from the Fermi level E$_{F}$ as a function of the compressed volume. The minus sign indicates that the point of interest is still below E$_{F}$ with a crossover at zero, after which it becomes positive. Compressing the volume, the nonsymmorphic band crossing approaches E$_{F}$ gradually, passes it at about $6$ $\%$ of volume compression, after which it continues to move away steadily. Upon application of a hydrostatic pressure, all initial symmetries in the system remain unaffected. The NN distances (d$_{1}$, d$_{2}$, d$_{3}$) decrease linearly with the increasing hydrostatic pressure. Accordingly, the whole crystal structure deforms smoothly under compression, yielding no signatures of deformations or electronic topological transitions. The two tiny, approximately spherical Fermi surfaces around $\Gamma$ point ($1$ and $2$) vanish with the application of the critical hydrostatic pressure of about 6 $\%$. \\
\hspace*{2em} To determine the bulk modulus and its derivative with respect to pressure, our calculated hydrostatic pressure data were fitted with the Birch-Murnaghan (BM) \cite{vinetequationofstate} equation of state (EOS). As a result of the BM fit, the bulk modulus and its pressure derivative are B $=$ $117$ GPa and B' $=$ $4.5$, respectively. In the literature, there are no experimental nor theoretical reports we can compare our results with. According to reference \cite{bulkmodulus}, which studies the elemental bulk modulus across the periodic table of elements, neither Hf nor Sn belong to the class of high bulk modulus materials. Hence, the value for HfSn is expected to be in the range typical for binary intermetallics \cite{bulkmodulusref}, in agreement with our findings. We note that fitting the data with Vinet \cite{vinetequationofstate}, Birch only and Murnaghan only \cite{murnaghanequationofstate} EOS result in a small renormalization of the bulk modulus (by maximum 5 GPa). Employing the volume-pressure dependence \cite{inversebmequation}, we estimate that $6\%$ of volume compression, required to locate the degeneracy at $E_{F}$, is equivalent to the application of hydrostatic pressure of about $8.5$ GPa. \\
\subsection{Uniaxial strain and chemical pressure}
\label{Uniaxialstrainandchemical pressure}
\begin{figure}[h!]
\includegraphics[width=0.47\textwidth, keepaspectratio]{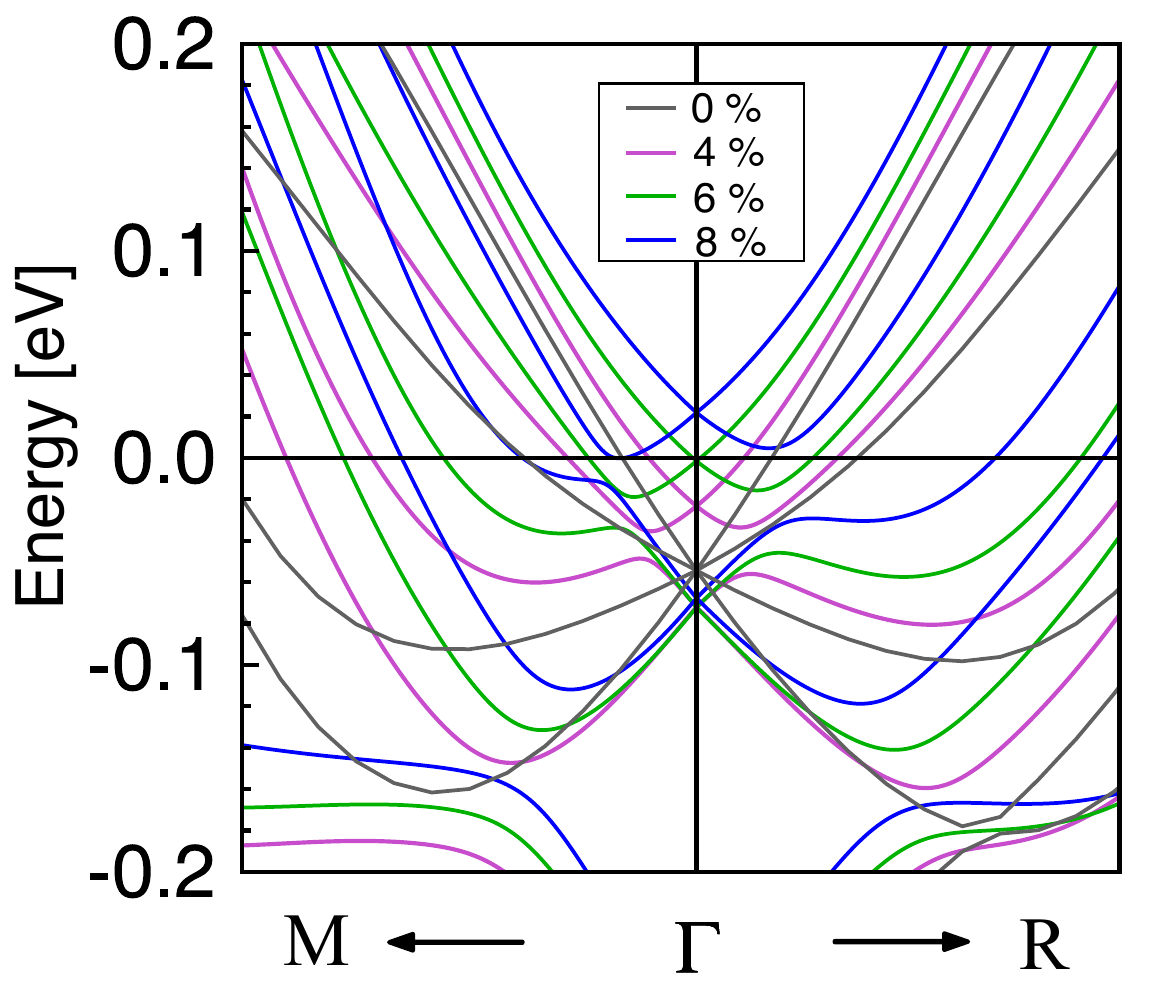}
\caption{Zoom of the bandstructure of HfSn around E$_{F}$ under strain. Zero strain structure depicted in grey is FPLO LDA optimized structure. }
\label{strainevolutionfigure}
\end{figure}

\hspace*{2em} However, the hydrostatic pressure is not the only path to pinpoint the Fermi level at the degenerate point. To achieve this, uniaxial strain and doping can be also utilized as tuning tools. Compared to the hydrostatic pressure, strain is expected to lead to much stronger effects influencing not just the position of the bands but also the band degeneracies due to the symmetry breaking \cite{barber}. First, a compressive strain along the $[100]$ lattice direction is applied, lifting the original cubic crystal symmetry. Application of strain along $[100]$ breaks the 3-fold rotational symmetry, all five simple and three screw operations, and consequently there is a lowering of the crystal symmetry from cubic to orthorombic. For the unstrained lattice parameters, our FPLO LDA optimized structure is exploited. Intrinsically, compression along the $a$ axis leads to a tension along $b$ and $c$ with specific, material dependent Poisson ratios. We assume an ideal Poisson ratio of $0.5$, often employed to describe linear elastic materials. As illustrated in Fig. $\ref{strainevolutionfigure}$, uniaxial compressive strain results in a splitting of the fourfold degenerate point at $\Gamma$ into two twofold degenerate points. At the R point, the native sixfold degenerate point with uniaxial strain transforms into three twofold crossings. The newly created pairs of twofold degeneracies at $\Gamma$ behave distinctly upon increasing strain. While one pair stays mostly in the vicinity of the original fourfold degenerate point, the second pair experiences a notable shift in energy and at about $6\%$ of strain it is situated at the Fermi level.   \\
\hspace*{2em} Creating off-stoichiometry at either Hf or Sn site is an alternative route to control the band filling, and hence the position of the bands with respect to the Fermi level. This manipulation induces chemical pressure which resembles the effect of hydrostatic pressure in a nonuniform manner. As a first step, we investigate this effect in a rigid band approximation. Assuming a simple band shift within LDA, it is required to add approximately 0.1 electrons per unit cell to displace the degenerate point at the Fermi level (in GGA, this point is even further away from the Fermi level, so approximately a three times larger number of electrons needs to be added to accomplish the same effect). To improve the description of a chemical disorder, as a next step we employ a virtual crystal approximation (VCA), which considers not only the shift in the Fermi energy, but also change in the potential traced to the modified number of electrons in the system \cite{vcabasic}. In practice, in the VCA approach the average charge at either Sn or Hf site is reduced.  We find that introducing a deficiency at the Sn site HfSn$_{x}$ of about $x=0.10$, the degenerate point is located at the Fermi level. This can be accomplished by substituting Sn atoms by Ge atoms (HfSn$_{1-x}$Ge$_{x}$). Both Sn and Ge are tetravalent atoms and have ionic radii which differ by almost $0.14$ $\si{\angstrom}$. Considering that Ge has smaller ionic radii compared to Sn, the volume of the unit cell is expected to decrease with this substitution, an effect which additionaly influences the position of the bands, as demonstrated previously. As both effects act in the same way, the actual percentage of Sn atoms to be substituted is expected to be smaller than $10$ $\%$. Equivalently, deficiency at the Hf site leads to a shift of the bands towards the Fermi level. The difference is that a slightly larger amount of Hf has to be substituted (at least $12$ $\%$ to achieve the same effect as $10$ $\%$ Sn substitution). Isovalent substitution of Hf with Zr is a possible mechanism as it does not introduce significant structural distortions due to the similarity of the ionic radii of Hf and Zr. \\
\subsection{Construction of an effective tight-binding model}
\label{tightbindingmodel}
\hspace*{2em} To complete the description of the compound, it is desirable to obtain a reliable effective tight-binding (TB) model that would not just reveal the main hopping mechanisms in the system but also enable access to the topological properties. To achieve this, we construct a tight-binding model based on maximally localized Wannier functions that reproduces accurately the band structure obtained by means of DFT (see Fig. $\ref{wannierfitwithoutSOC}$ and $\ref{wannierfitwithSOC}$ in Appendix \ref{Wannierfit}). As initial projectors for constructing the Wannier Hamiltonian, we employ Hf $5$d ($xy$, $yz$, $xz$, $x^{2}$-$y^{2}$, $z^{2}$), $6$p ($x$, $y$, $z$) and Sn $5$p ($x$, $y$, $z$) orbitals.  Due to the fact that the target bands in the vicinity of the Fermi level, essential for construction of the low-energy model, are strongly hybridized, one needs to set a large energy window while constructing the WFs. In addition, we have carefully examined the dependence of the results on the real space distance at which the hoppings are cut off and the magnitude of the hoppings with respect to the specified threshold (both requirements ensure that the minimum number of hoppings, essential for the physics, are contained in the model). Both have to be large in order to reliably reproduce the DFT band structure. Due to the lack of inversion symmetry, all hopping parameters along arbitrary directions $\mathbf{r}$ and $-\mathbf{r}$ are allowed to differ in magnitude and sign. Our calculations reveal that nearest neighbour (NN) coupling between Hf and Sn atoms is the dominant hopping regime in the compound. It is driven by a strong hybridization between Hf $5$d and Sn $5$p orbitals, as well as Hf $6$p and Sn $5$p states (see Tables \ref{tableofnearestneighbourhoppingparameters} and \ref{tableofnextnearestneighbourhoppingparameters} in the Appendix \ref{tableswithandwithoutsoc}). The hopping amplitudes from Hf $5$d to Sn $5$p states are t$_{\text{pd}}$ $\approx$ $1.2$ eV (one out of seven possible coupling mechanisms). Interestingly, hopping to the Sn $5$p states from Hf $6$p state is even larger compared to the d-p channel, and these are the parameters with the largest magnitude ($\approx$ $1.7$ eV). This is elucidated by the fact that the Hf $6$p orbitals are extended enough to provide significant overlap with the neighbouring p orbitals of Sn atoms. Moreover, the overlap between $5$p and $6$p orbitals of the same kind (p$_{x}$-p$_{x}$, p$_{y}$-p$_{y}$, p$_{z}$-p$_{z}$) is by a factor of two larger than the overlap between orbitals of different kind. All the next nearest neighbour (NNN) hopping parameters are decreased compared to the NN ones, in agreement with the fact that the hopping magnitude decreases with increasing distance between the sites. It should be emphasised that NNN parameters cannot be neglected when building the effective model, signalising their important role in reliable compound description. Quite surprisingly, the long range hopping parameters are of enormous importance in obtaining a reliable fit to the DFT band structure. The so derived picture of the hopping regime of HfSn remains to hold with the inclusion of SOC, just with a renormalization of some of the parameters and introduction of imaginary components (see Tables \ref{tableofnearestneighbourhoppingparameterssoc} and \ref{tableofnextnearestneighbourhoppingparameterssoc} in the Appendix \ref{tableswithandwithoutsoc}). In particular, the p-p overlaps in the NN coordination polyhedra are affected. This could be attributed to the relativistic effects which have tendency to contract the p orbitals to some extent.     \\
\hspace*{2em} Using a TB model Hamiltonian, we evaluate the band-resolved Berry curvature (BC) and the BC summed over the occupied states depicted in (a) and (b) of Fig. $\ref{berry}$ (Appendix \ref{berrycurvature}), respectively. Our calculations illustrate that there is a large contribution to the Berry curvature when bands are near each other energetically. As the next step, the topological charges of the bands forming the fourfold degenerate point at the $\Gamma$ point just below the E$_{F}$ are evaluated. The dependence of the results on the radius of a sphere constructed around $\Gamma$ is carefully examined. We find that the topological charges for the 1th-4th bands are +3 (band 1), +1 (band 2), -1 (band 3) and -3 (band 4) (see notation in Fig. \ref{bulkbandstructure} (b)). The values for the topological charges and their distribution across bands agree well with those derived for CoGe \cite{coge} and CoSi \cite{cosiseverin} for their fourfold degenerate points located approximately $20$ meV below E$_{F}$ for both systems. The topological character of the fourfold degeneracy point identifies this point as a spin-$3/2$ Weyl node, in agreement with \cite{coge}. Multifold fermions at high-symmetry points, in particular fourfold spin-$3/2$ chiral Weyl fermions at $\Gamma$ and sixfold Weyl fermions at R, are characteristic for B$20$ compounds and are expected to give rise to unique topology-related transport properties \cite{orbitaltexture} \cite{opticalresponse} \cite{quantized}, if they are close to E$_{F}$.     
\section{Summary and Conclusion}
\hspace*{2em} In summary, we perform an extensive theoretical study of noncentrosymmetric HfSn, first known member of the B$20$ family where a transition metal is from the fourth group. We investigate the band structure and Fermi surface utilizing DFT and an effective model approach to access the topological properties. Our results presented above show that there is a fourfold degenerate point at $\Gamma$ located maximum $0.1$ eV below the Fermi level, which has spin-$3/2$ Weyl node character. Considering that this band energy is not tremendously large, it is possible to tune it across the Fermi level via external perturbations. Such a manipulation would conceivably lead to a spectrum of novel physical phenomena like for e.g. chiral anomaly \cite{anomaly} and large nonlinear optical conductivity \cite{opticalpaper}. We demonstrate that the volume shrinkage caused by increasing hydrostatic pressure is driving the fourfold degeneracy towards the Fermi level, reaches it for about $6$$\%$ of compression, and continues to relocate in a same manner. In contrast to hydrostatic pressure where the volume is compressed isotropically, the uniaxial strain allows selection of specific lattice direction enabling the control over breaking of the original symmetries of the crystal lattice. Applying the $[100]$ strain on HfSn, we lift some symmetries and break fourfold degeneracy into two twofold subsets, one of which is driven across the Fermi level for strain of about $6\%$. Comparing pressure and strain, the hydrostatic pressure investigations are at present more applicable to HfSn due to the availability of powder samples only. However, constant advancements of the synthesis techniques may result in high quality single crystals of HfSn in near future, highlighting the significance of the uniaxial pressure study presented here. Creating off-stoichiometry has a considerable effect on the electronic structure of HfSn. The isoelectronic substitution of Sn by Ge shifts the Fermi level, and fourfold point is at $\Gamma$ for Ge concentration of about $10\%$. Further investigations are required to understand how the disorder created by this perturbation influences other bands. Nevertheless, neither hydrostatic pressure nor chemical composition of the sample have influence on the multiplicity of the band crossings occurring at high symmetry points. They arise purely due to the symmetry elements of the respective chiral space group. The crucial factor capable of modifying this multiplicity is the symmetry breaking that can be realized by the application of uniaxial strain. All three ways of external tuning presented in the current work move the crossing up and down in energy, but only strain prompts a reduction of its degeneracy. Parametrization of an effective TB model reveals the dominant role of NN and NNN hoppings with nonnegligible contribution of the higher-order terms. Although it remains to see whether this conclusion is generic across the whole family of B$20$ compounds, a consistent set of parameters derived in the current work undoubtedly can be utilized as a starting point to fit the band structure of a different system in the same class of compounds. This considerably speeds up the construction of an effective model.     \\
\hspace*{2em} The methodology presented in the current work is not specific to HfSn only, but can be equally applied to other members of the B$20$ family. It is particularly relevant to those compounds which have their symmetry-protected crossings within few meV range from the Fermi level. Applying exactly the same approach, one obtains initial predictions about binding energies of the respective bands and energy window within which they can be shifted by different mechanisms of external perturbations. Our results demonstrate how DFT can be employed as a useful tool to qualitatively predict the outcome of complex experiments before they are performed and can help designing new experimental setups. Going beyond B$20$, many other space groups host symmetry protected band crossings which can be approached in the same way as presented here, illustrating the generality of our findings.
\section*{Acknowledgements}
Computations were performed on the HPC systems Raven and Cobra at the Max Planck Computing and Data Facility (MP CDF). D.M. acknowledges application support provided by MP CDF.
\nocite{*}

\begin{thebibliography}{46}%
\makeatletter
\providecommand \@ifxundefined [1]{%
 \@ifx{#1\undefined}
}%
\providecommand \@ifnum [1]{%
 \ifnum #1\expandafter \@firstoftwo
 \else \expandafter \@secondoftwo
 \fi
}%
\providecommand \@ifx [1]{%
 \ifx #1\expandafter \@firstoftwo
 \else \expandafter \@secondoftwo
 \fi
}%
\providecommand \natexlab [1]{#1}%
\providecommand \enquote  [1]{``#1''}%
\providecommand \bibnamefont  [1]{#1}%
\providecommand \bibfnamefont [1]{#1}%
\providecommand \citenamefont [1]{#1}%
\providecommand \href@noop [0]{\@secondoftwo}%
\providecommand \href [0]{\begingroup \@sanitize@url \@href}%
\providecommand \@href[1]{\@@startlink{#1}\@@href}%
\providecommand \@@href[1]{\endgroup#1\@@endlink}%
\providecommand \@sanitize@url [0]{\catcode `\\12\catcode `\$12\catcode
  `\&12\catcode `\#12\catcode `\^12\catcode `\_12\catcode `\%12\relax}%
\providecommand \@@startlink[1]{}%
\providecommand \@@endlink[0]{}%
\providecommand \url  [0]{\begingroup\@sanitize@url \@url }%
\providecommand \@url [1]{\endgroup\@href {#1}{\urlprefix }}%
\providecommand \urlprefix  [0]{URL }%
\providecommand \Eprint [0]{\href }%
\providecommand \doibase [0]{http://dx.doi.org/}%
\providecommand \selectlanguage [0]{\@gobble}%
\providecommand \bibinfo  [0]{\@secondoftwo}%
\providecommand \bibfield  [0]{\@secondoftwo}%
\providecommand \translation [1]{[#1]}%
\providecommand \BibitemOpen [0]{}%
\providecommand \bibitemStop [0]{}%
\providecommand \bibitemNoStop [0]{.\EOS\space}%
\providecommand \EOS [0]{\spacefactor3000\relax}%
\providecommand \BibitemShut  [1]{\csname bibitem#1\endcsname}%
\let\auto@bib@innerbib\@empty
\bibitem [{\citenamefont {Fecher}\ \emph {et~al.}(2022)\citenamefont {Fecher},
  \citenamefont {K{\"u}bler},\ and\ \citenamefont {Felser}}]{fecher}%
  \BibitemOpen
  \bibfield  {author} {\bibinfo {author} {\bibfnamefont {G.~H.}\ \bibnamefont
  {Fecher}}, \bibinfo {author} {\bibfnamefont {J.}~\bibnamefont {K{\"u}bler}},
  \ and\ \bibinfo {author} {\bibfnamefont {C.}~\bibnamefont {Felser}},\
  }\bibfield  {title} {\enquote {\bibinfo {title} {Chirality in the solid
  state: {C}hiral {C}rystal {S}tructures in {C}hiral and {A}chiral {S}pace
  {G}roups},}\ }\href@noop {} {\bibfield  {journal} {\bibinfo  {journal}
  {Materials}\ }\textbf {\bibinfo {volume} {15}},\ \bibinfo {pages} {5812}
  (\bibinfo {year} {2022})}\BibitemShut {NoStop}%
\bibitem [{\citenamefont {Franciosi}\ \emph {et~al.}(1982)\citenamefont
  {Franciosi}, \citenamefont {Peterman}, \citenamefont {Weaver},\ and\
  \citenamefont {Moruzzi}}]{CrSimetal}%
  \BibitemOpen
  \bibfield  {author} {\bibinfo {author} {\bibfnamefont {A.}~\bibnamefont
  {Franciosi}}, \bibinfo {author} {\bibfnamefont {D.~J.}\ \bibnamefont
  {Peterman}}, \bibinfo {author} {\bibfnamefont {J.~H.}\ \bibnamefont
  {Weaver}}, \ and\ \bibinfo {author} {\bibfnamefont {V.~L.}\ \bibnamefont
  {Moruzzi}},\ }\bibfield  {title} {\enquote {\bibinfo {title} {Structural
  morphology and electronic properties of the {S}i-{C}r interface},}\
  }\href@noop {} {\bibfield  {journal} {\bibinfo  {journal} {Phys. Rev. B}\
  }\textbf {\bibinfo {volume} {25}},\ \bibinfo {pages} {4981--4993} (\bibinfo
  {year} {1982})}\BibitemShut {NoStop}%
\bibitem [{\citenamefont {Vescoli}\ \emph {et~al.}(1998)\citenamefont
  {Vescoli}, \citenamefont {Degiorgi}, \citenamefont {Buschinger},
  \citenamefont {Guth}, \citenamefont {Geibel},\ and\ \citenamefont
  {Steglich}}]{kondoinsulator}%
  \BibitemOpen
  \bibfield  {author} {\bibinfo {author} {\bibfnamefont {V.}~\bibnamefont
  {Vescoli}}, \bibinfo {author} {\bibfnamefont {L.}~\bibnamefont {Degiorgi}},
  \bibinfo {author} {\bibfnamefont {B.}~\bibnamefont {Buschinger}}, \bibinfo
  {author} {\bibfnamefont {W.}~\bibnamefont {Guth}}, \bibinfo {author}
  {\bibfnamefont {C.}~\bibnamefont {Geibel}}, \ and\ \bibinfo {author}
  {\bibfnamefont {F.}~\bibnamefont {Steglich}},\ }\bibfield  {title} {\enquote
  {\bibinfo {title} {The optical properties of {R}u{S}i: Kondo insulator or
  conventional semiconductor?}}\ }\href@noop {} {\bibfield  {journal} {\bibinfo
   {journal} {Solid State Communications}\ }\textbf {\bibinfo {volume} {105}},\
  \bibinfo {pages} {367--370} (\bibinfo {year} {1998})}\BibitemShut {NoStop}%
\bibitem [{\citenamefont {Datta}\ \emph {et~al.}(2022)\citenamefont {Datta},
  \citenamefont {Vasdev}, \citenamefont {Rana}, \citenamefont {Motla},
  \citenamefont {Kataria}, \citenamefont {Singh}, \citenamefont {Das},\ and\
  \citenamefont {Sheet}}]{aubesuperconductor}%
  \BibitemOpen
  \bibfield  {author} {\bibinfo {author} {\bibfnamefont {S.}~\bibnamefont
  {Datta}}, \bibinfo {author} {\bibfnamefont {A.}~\bibnamefont {Vasdev}},
  \bibinfo {author} {\bibfnamefont {P.~S.}\ \bibnamefont {Rana}}, \bibinfo
  {author} {\bibfnamefont {K.}~\bibnamefont {Motla}}, \bibinfo {author}
  {\bibfnamefont {A.}~\bibnamefont {Kataria}}, \bibinfo {author} {\bibfnamefont
  {R.~P.}\ \bibnamefont {Singh}}, \bibinfo {author} {\bibfnamefont
  {T.}~\bibnamefont {Das}}, \ and\ \bibinfo {author} {\bibfnamefont
  {G.}~\bibnamefont {Sheet}},\ }\bibfield  {title} {\enquote {\bibinfo {title}
  {Spectroscopic evidence of multigap superconductivity in noncentrosymmetric
  {A}u{B}e},}\ }\href@noop {} {\bibfield  {journal} {\bibinfo  {journal} {Phys.
  Rev. B}\ }\textbf {\bibinfo {volume} {105}},\ \bibinfo {pages} {104505}
  (\bibinfo {year} {2022})}\BibitemShut {NoStop}%
\bibitem [{\citenamefont {M{\"u}hlbauer}\ \emph {et~al.}(2009)\citenamefont
  {M{\"u}hlbauer}, \citenamefont {Binz}, \citenamefont {Jonietz}, \citenamefont
  {Pfleiderer}, \citenamefont {Rosch}, \citenamefont {Neubauer}, \citenamefont
  {Georgii},\ and\ \citenamefont {B{\"o}ni}}]{mnsiskyrmions}%
  \BibitemOpen
  \bibfield  {author} {\bibinfo {author} {\bibfnamefont {S.}~\bibnamefont
  {M{\"u}hlbauer}}, \bibinfo {author} {\bibfnamefont {B.}~\bibnamefont {Binz}},
  \bibinfo {author} {\bibfnamefont {F.}~\bibnamefont {Jonietz}}, \bibinfo
  {author} {\bibfnamefont {C.}~\bibnamefont {Pfleiderer}}, \bibinfo {author}
  {\bibfnamefont {A.}~\bibnamefont {Rosch}}, \bibinfo {author} {\bibfnamefont
  {A.}~\bibnamefont {Neubauer}}, \bibinfo {author} {\bibfnamefont
  {R.}~\bibnamefont {Georgii}}, \ and\ \bibinfo {author} {\bibfnamefont
  {P.}~\bibnamefont {B{\"o}ni}},\ }\bibfield  {title} {\enquote {\bibinfo
  {title} {Skyrmion lattice in a {C}hiral magnet},}\ }\href@noop {} {\bibfield
  {journal} {\bibinfo  {journal} {Science}\ }\textbf {\bibinfo {volume}
  {323}},\ \bibinfo {pages} {915--919} (\bibinfo {year} {2009})}\BibitemShut
  {NoStop}%
\bibitem [{\citenamefont {Schr{\"o}ter}\ \emph {et~al.}(2020)\citenamefont
  {Schr{\"o}ter}, \citenamefont {Stolz}, \citenamefont {Manna}, \citenamefont
  {de~Juan}, \citenamefont {Vergniory}, \citenamefont {Krieger}, \citenamefont
  {Pei}, \citenamefont {Schmitt}, \citenamefont {Dudin}, \citenamefont {Kim},
  \citenamefont {Cacho}, \citenamefont {Bradlyn}, \citenamefont {Borrmann},
  \citenamefont {Schmidt}, \citenamefont {Widmer}, \citenamefont {Strocov},\
  and\ \citenamefont {Felser}}]{maximalchernnumber}%
  \BibitemOpen
  \bibfield  {author} {\bibinfo {author} {\bibfnamefont {N.~B.~M.}\
  \bibnamefont {Schr{\"o}ter}}, \bibinfo {author} {\bibfnamefont
  {S.}~\bibnamefont {Stolz}}, \bibinfo {author} {\bibfnamefont
  {K.}~\bibnamefont {Manna}}, \bibinfo {author} {\bibfnamefont
  {F.}~\bibnamefont {de~Juan}}, \bibinfo {author} {\bibfnamefont {M.~G.}\
  \bibnamefont {Vergniory}}, \bibinfo {author} {\bibfnamefont {J.~A.}\
  \bibnamefont {Krieger}}, \bibinfo {author} {\bibfnamefont {D.}~\bibnamefont
  {Pei}}, \bibinfo {author} {\bibfnamefont {T.}~\bibnamefont {Schmitt}},
  \bibinfo {author} {\bibfnamefont {P.}~\bibnamefont {Dudin}}, \bibinfo
  {author} {\bibfnamefont {T.~K.}\ \bibnamefont {Kim}}, \bibinfo {author}
  {\bibfnamefont {C.}~\bibnamefont {Cacho}}, \bibinfo {author} {\bibfnamefont
  {B.}~\bibnamefont {Bradlyn}}, \bibinfo {author} {\bibfnamefont
  {H.}~\bibnamefont {Borrmann}}, \bibinfo {author} {\bibfnamefont
  {M.}~\bibnamefont {Schmidt}}, \bibinfo {author} {\bibfnamefont
  {R.}~\bibnamefont {Widmer}}, \bibinfo {author} {\bibfnamefont {V.~N.}\
  \bibnamefont {Strocov}}, \ and\ \bibinfo {author} {\bibfnamefont
  {C.}~\bibnamefont {Felser}},\ }\bibfield  {title} {\enquote {\bibinfo {title}
  {Observation and control of maximal {C}hern numbers in a chiral topological
  semimetal},}\ }\href@noop {} {\bibfield  {journal} {\bibinfo  {journal}
  {Science}\ }\textbf {\bibinfo {volume} {369}},\ \bibinfo {pages} {179--183}
  (\bibinfo {year} {2020})}\BibitemShut {NoStop}%
\bibitem [{\citenamefont {Yang}\ \emph {et~al.}(2020)\citenamefont {Yang},
  \citenamefont {Li}, \citenamefont {Manna}, \citenamefont {Fan}, \citenamefont
  {Felser},\ and\ \citenamefont {Sun}}]{qunyang}%
  \BibitemOpen
  \bibfield  {author} {\bibinfo {author} {\bibfnamefont {Q.}~\bibnamefont
  {Yang}}, \bibinfo {author} {\bibfnamefont {G.}~\bibnamefont {Li}}, \bibinfo
  {author} {\bibfnamefont {K.}~\bibnamefont {Manna}}, \bibinfo {author}
  {\bibfnamefont {F.}~\bibnamefont {Fan}}, \bibinfo {author} {\bibfnamefont
  {C.}~\bibnamefont {Felser}}, \ and\ \bibinfo {author} {\bibfnamefont
  {Y.}~\bibnamefont {Sun}},\ }\bibfield  {title} {\enquote {\bibinfo {title}
  {Topological {E}ngineering of {P}t-{G}roup-{M}etal-{B}ased {C}hiral
  {C}rystals toward {H}igh-{E}fficiency {H}ydrogen {E}volution {C}atalysts},}\
  }\href {\doibase https://doi.org/10.1002/adma.201908518} {\bibfield
  {journal} {\bibinfo  {journal} {Advanced Materials}\ }\textbf {\bibinfo
  {volume} {32}},\ \bibinfo {pages} {1908518} (\bibinfo {year}
  {2020})}\BibitemShut {NoStop}%
\bibitem [{\citenamefont {Pshenay-Severin}\ and\ \citenamefont
  {Burkov}(2019)}]{severin}%
  \BibitemOpen
  \bibfield  {author} {\bibinfo {author} {\bibfnamefont {D.~A.}\ \bibnamefont
  {Pshenay-Severin}}\ and\ \bibinfo {author} {\bibfnamefont {A.~T.}\
  \bibnamefont {Burkov}},\ }\bibfield  {title} {\enquote {\bibinfo {title}
  {Electronic {S}tructure of {B20} ({F}e{S}i-type) {T}ransition-{M}etal
  {M}onosilicides},}\ }\href@noop {} {\bibfield  {journal} {\bibinfo  {journal}
  {Materials}\ }\textbf {\bibinfo {volume} {12}},\ \bibinfo {pages} {2710}
  (\bibinfo {year} {2019})}\BibitemShut {NoStop}%
\bibitem [{\citenamefont {Schob}\ and\ \citenamefont
  {Parth{\'{e}}}(1964)}]{schob}%
  \BibitemOpen
  \bibfield  {author} {\bibinfo {author} {\bibfnamefont {O.}~\bibnamefont
  {Schob}}\ and\ \bibinfo {author} {\bibfnamefont {E.}~\bibnamefont
  {Parth{\'{e}}}},\ }\bibfield  {title} {\enquote {\bibinfo {title} {{The
  structure of HfSn}},}\ }\href@noop {} {\bibfield  {journal} {\bibinfo
  {journal} {Acta Crystallographica}\ }\textbf {\bibinfo {volume} {17}},\
  \bibinfo {pages} {452--453} (\bibinfo {year} {1964})}\BibitemShut {NoStop}%
\bibitem [{\citenamefont {Hong}\ and\ \citenamefont {van~de
  Walle}(2015)}]{highestmeltingpoint}%
  \BibitemOpen
  \bibfield  {author} {\bibinfo {author} {\bibfnamefont {Q.-J.}\ \bibnamefont
  {Hong}}\ and\ \bibinfo {author} {\bibfnamefont {A.}~\bibnamefont {van~de
  Walle}},\ }\bibfield  {title} {\enquote {\bibinfo {title} {Prediction of the
  material with highest known melting point from ab initio molecular dynamics
  calculations},}\ }\href@noop {} {\bibfield  {journal} {\bibinfo  {journal}
  {Phys. Rev. B}\ }\textbf {\bibinfo {volume} {92}},\ \bibinfo {pages} {020104}
  (\bibinfo {year} {2015})}\BibitemShut {NoStop}%
\bibitem [{\citenamefont {Zhou}\ \emph {et~al.}(2016)\citenamefont {Zhou},
  \citenamefont {Shao}, \citenamefont {Shi}, \citenamefont {Sun}, \citenamefont
  {Felser}, \citenamefont {Yan},\ and\ \citenamefont {Frauenheim}}]{strain}%
  \BibitemOpen
  \bibfield  {author} {\bibinfo {author} {\bibfnamefont {L.}~\bibnamefont
  {Zhou}}, \bibinfo {author} {\bibfnamefont {B.}~\bibnamefont {Shao}}, \bibinfo
  {author} {\bibfnamefont {W.}~\bibnamefont {Shi}}, \bibinfo {author}
  {\bibfnamefont {Y.}~\bibnamefont {Sun}}, \bibinfo {author} {\bibfnamefont
  {C.}~\bibnamefont {Felser}}, \bibinfo {author} {\bibfnamefont
  {B.}~\bibnamefont {Yan}}, \ and\ \bibinfo {author} {\bibfnamefont
  {T.}~\bibnamefont {Frauenheim}},\ }\bibfield  {title} {\enquote {\bibinfo
  {title} {Prediction of the quantum spin {H}all effect in monolayers of
  transition-metal carbides {MC} ({M = Ti, Zr, Hf})},}\ }\href@noop {}
  {\bibfield  {journal} {\bibinfo  {journal} {2D Materials}\ }\textbf {\bibinfo
  {volume} {3}},\ \bibinfo {pages} {035022} (\bibinfo {year}
  {2016})}\BibitemShut {NoStop}%
\bibitem [{\citenamefont {Gao}\ \emph {et~al.}(2021)\citenamefont {Gao},
  \citenamefont {Iitaka},\ and\ \citenamefont {Li}}]{opticalpaper}%
  \BibitemOpen
  \bibfield  {author} {\bibinfo {author} {\bibfnamefont {Y.}~\bibnamefont
  {Gao}}, \bibinfo {author} {\bibfnamefont {T.}~\bibnamefont {Iitaka}}, \ and\
  \bibinfo {author} {\bibfnamefont {Z.}~\bibnamefont {Li}},\ }\bibfield
  {title} {\enquote {\bibinfo {title} {Terahertz nonlinear optics of chiral
  semimetals {RhSn}, {HfSn}, and {PdGa}},}\ }\href@noop {} {\bibfield
  {journal} {\bibinfo  {journal} {The European Physical Journal B}\ }\textbf
  {\bibinfo {volume} {94}},\ \bibinfo {pages} {95} (\bibinfo {year}
  {2021})}\BibitemShut {NoStop}%
\bibitem [{\citenamefont {Huang}\ \emph {et~al.}(2018)\citenamefont {Huang},
  \citenamefont {Chen}, \citenamefont {Macam}, \citenamefont {Crisostomo},
  \citenamefont {Huang}, \citenamefont {Chen}, \citenamefont {Albao},
  \citenamefont {Jang}, \citenamefont {Lin},\ and\ \citenamefont
  {Chuang}}]{huangnanoscale}%
  \BibitemOpen
  \bibfield  {author} {\bibinfo {author} {\bibfnamefont {Z.-Q.}\ \bibnamefont
  {Huang}}, \bibinfo {author} {\bibfnamefont {W.-C.}\ \bibnamefont {Chen}},
  \bibinfo {author} {\bibfnamefont {G.}~\bibnamefont {Macam}}, \bibinfo
  {author} {\bibfnamefont {C.}~\bibnamefont {Crisostomo}}, \bibinfo {author}
  {\bibfnamefont {S.-M.}\ \bibnamefont {Huang}}, \bibinfo {author}
  {\bibfnamefont {R.-B.}\ \bibnamefont {Chen}}, \bibinfo {author}
  {\bibfnamefont {M.}~\bibnamefont {Albao}}, \bibinfo {author} {\bibfnamefont
  {D.-J.}\ \bibnamefont {Jang}}, \bibinfo {author} {\bibfnamefont
  {H.}~\bibnamefont {Lin}}, \ and\ \bibinfo {author} {\bibfnamefont {F.-C.}\
  \bibnamefont {Chuang}},\ }\bibfield  {title} {\enquote {\bibinfo {title}
  {Predicition of {Q}uantum {A}nomalous {H}all effect in {MB}i and {MS}b
  ({M:Ti}, {Zr}, and {Hf}) {H}oneycombs},}\ }\href@noop {} {\bibfield
  {journal} {\bibinfo  {journal} {Nanoscale Research Letters}\ }\textbf
  {\bibinfo {volume} {13}},\ \bibinfo {pages} {43} (\bibinfo {year}
  {2018})}\BibitemShut {NoStop}%
\bibitem [{\citenamefont {Tsyganova}\ \emph {et~al.}(1971)\citenamefont
  {Tsyganova}, \citenamefont {Tylkina},\ and\ \citenamefont
  {Savitskiy}}]{tsyganova}%
  \BibitemOpen
  \bibfield  {author} {\bibinfo {author} {\bibfnamefont {I.}~\bibnamefont
  {Tsyganova}}, \bibinfo {author} {\bibfnamefont {M.}~\bibnamefont {Tylkina}},
  \ and\ \bibinfo {author} {\bibfnamefont {E.}~\bibnamefont {Savitskiy}},\
  }\bibfield  {title} {\enquote {\bibinfo {title} {{Sn-Hf} and {Sn-Re} phase
  diagrams},}\ }\href@noop {} {\bibfield  {journal} {\bibinfo  {journal} {Izv.
  Akad. Nauk SSSR, Met.,}\ }\textbf {\bibinfo {volume} {3}},\ \bibinfo {pages}
  {188--191} (\bibinfo {year} {1971})}\BibitemShut {NoStop}%
\bibitem [{\citenamefont {Romaka}\ \emph {et~al.}(2001)\citenamefont {Romaka},
  \citenamefont {Stadnyk},\ and\ \citenamefont {Bodak}}]{romaka}%
  \BibitemOpen
  \bibfield  {author} {\bibinfo {author} {\bibfnamefont {L.}~\bibnamefont
  {Romaka}}, \bibinfo {author} {\bibfnamefont {Y.}~\bibnamefont {Stadnyk}}, \
  and\ \bibinfo {author} {\bibfnamefont {O.}~\bibnamefont {Bodak}},\ }\bibfield
   {title} {\enquote {\bibinfo {title} {Ternary {Hf-Co-Sn} system},}\
  }\href@noop {} {\bibfield  {journal} {\bibinfo  {journal} {Journal of Alloys
  and Compounds}\ }\textbf {\bibinfo {volume} {317-318}},\ \bibinfo {pages}
  {347--349} (\bibinfo {year} {2001})}\BibitemShut {NoStop}%
\bibitem [{\citenamefont {Stadnyk}\ and\ \citenamefont
  {Romaka}(2001)}]{stadnyak}%
  \BibitemOpen
  \bibfield  {author} {\bibinfo {author} {\bibfnamefont {Y.}~\bibnamefont
  {Stadnyk}}\ and\ \bibinfo {author} {\bibfnamefont {L.}~\bibnamefont
  {Romaka}},\ }\bibfield  {title} {\enquote {\bibinfo {title} {Phase equilibria
  in the {Hf--Ni--Sn} ternary system and crystal structures of the
  {Hf}2{Ni}2{Sn} compound},}\ }\href@noop {} {\bibfield  {journal} {\bibinfo
  {journal} {Journal of Alloys and Compounds}\ }\textbf {\bibinfo {volume}
  {316}},\ \bibinfo {pages} {169--171} (\bibinfo {year} {2001})}\BibitemShut
  {NoStop}%
\bibitem [{\citenamefont {Tang}\ \emph {et~al.}(2012)\citenamefont {Tang},
  \citenamefont {Hu}, \citenamefont {Du}, \citenamefont {Zhao}, \citenamefont
  {Zhou}, \citenamefont {Zheng}, \citenamefont {Gao},\ and\ \citenamefont
  {Wang}}]{tang}%
  \BibitemOpen
  \bibfield  {author} {\bibinfo {author} {\bibfnamefont {C.}~\bibnamefont
  {Tang}}, \bibinfo {author} {\bibfnamefont {B.}~\bibnamefont {Hu}}, \bibinfo
  {author} {\bibfnamefont {Y.}~\bibnamefont {Du}}, \bibinfo {author}
  {\bibfnamefont {D.}~\bibnamefont {Zhao}}, \bibinfo {author} {\bibfnamefont
  {P.}~\bibnamefont {Zhou}}, \bibinfo {author} {\bibfnamefont {F.}~\bibnamefont
  {Zheng}}, \bibinfo {author} {\bibfnamefont {Q.}~\bibnamefont {Gao}}, \ and\
  \bibinfo {author} {\bibfnamefont {J.}~\bibnamefont {Wang}},\ }\bibfield
  {title} {\enquote {\bibinfo {title} {Thermodynamic modeling of the {Hf--Sn}
  and {Sn--Y} systems},}\ }\href@noop {} {\bibfield  {journal} {\bibinfo
  {journal} {Elsevier CALPHAD}\ }\textbf {\bibinfo {volume} {39}},\ \bibinfo
  {pages} {91--96} (\bibinfo {year} {2012})}\BibitemShut {NoStop}%
\bibitem [{\citenamefont {Koepernik}\ and\ \citenamefont
  {Eschrig}(1999)}]{fplo}%
  \BibitemOpen
  \bibfield  {author} {\bibinfo {author} {\bibfnamefont {K.}~\bibnamefont
  {Koepernik}}\ and\ \bibinfo {author} {\bibfnamefont {H.}~\bibnamefont
  {Eschrig}},\ }\bibfield  {title} {\enquote {\bibinfo {title} {Full-potential
  nonorthogonal local-orbital minimum-basis band-structure scheme},}\
  }\href@noop {} {\bibfield  {journal} {\bibinfo  {journal} {Phys. Rev. B}\
  }\textbf {\bibinfo {volume} {59}},\ \bibinfo {pages} {1743--1757} (\bibinfo
  {year} {1999})}\BibitemShut {NoStop}%
\bibitem [{\citenamefont {Kresse}\ and\ \citenamefont
  {Hafner}(1993)}]{vaspcitaton1}%
  \BibitemOpen
  \bibfield  {author} {\bibinfo {author} {\bibfnamefont {G.}~\bibnamefont
  {Kresse}}\ and\ \bibinfo {author} {\bibfnamefont {J.}~\bibnamefont
  {Hafner}},\ }\bibfield  {title} {\enquote {\bibinfo {title} {Ab initio
  molecular dynamics for liquid metals},}\ }\href@noop {} {\bibfield  {journal}
  {\bibinfo  {journal} {Phys. Rev. B}\ }\textbf {\bibinfo {volume} {47}},\
  \bibinfo {pages} {558--561} (\bibinfo {year} {1993})}\BibitemShut {NoStop}%
\bibitem [{\citenamefont {Kresse}\ and\ \citenamefont
  {Furthm{\"u}ller}(1996{\natexlab{a}})}]{vaspcitation2}%
  \BibitemOpen
  \bibfield  {author} {\bibinfo {author} {\bibfnamefont {G.}~\bibnamefont
  {Kresse}}\ and\ \bibinfo {author} {\bibfnamefont {J.}~\bibnamefont
  {Furthm{\"u}ller}},\ }\bibfield  {title} {\enquote {\bibinfo {title}
  {Efficiency of ab-initio total energy calculations for metals and
  semiconductors using a plane-wave basis set},}\ }\href@noop {} {\bibfield
  {journal} {\bibinfo  {journal} {Computational Material Science}\ }\textbf
  {\bibinfo {volume} {6}},\ \bibinfo {pages} {15--50} (\bibinfo {year}
  {1996}{\natexlab{a}})}\BibitemShut {NoStop}%
\bibitem [{\citenamefont {Kresse}\ and\ \citenamefont
  {Furthm{\"u}ller}(1996{\natexlab{b}})}]{vaspcitation3}%
  \BibitemOpen
  \bibfield  {author} {\bibinfo {author} {\bibfnamefont {G.}~\bibnamefont
  {Kresse}}\ and\ \bibinfo {author} {\bibfnamefont {J.}~\bibnamefont
  {Furthm{\"u}ller}},\ }\bibfield  {title} {\enquote {\bibinfo {title}
  {Efficient iterative schemes for ab initio total-energy calculations using a
  plane-wave basis set},}\ }\href@noop {} {\bibfield  {journal} {\bibinfo
  {journal} {Phys. Rev. B}\ }\textbf {\bibinfo {volume} {54}},\ \bibinfo
  {pages} {11169--11186} (\bibinfo {year} {1996}{\natexlab{b}})}\BibitemShut
  {NoStop}%
\bibitem [{\citenamefont {Perdew}\ and\ \citenamefont
  {Wang}(1992)}]{perdewwang}%
  \BibitemOpen
  \bibfield  {author} {\bibinfo {author} {\bibfnamefont {J.~P.}\ \bibnamefont
  {Perdew}}\ and\ \bibinfo {author} {\bibfnamefont {Y.}~\bibnamefont {Wang}},\
  }\bibfield  {title} {\enquote {\bibinfo {title} {Accurate and simple analytic
  representation of the electron-gas correlation energy},}\ }\href@noop {}
  {\bibfield  {journal} {\bibinfo  {journal} {Phys. Rev. B}\ }\textbf {\bibinfo
  {volume} {45}},\ \bibinfo {pages} {13244--13249} (\bibinfo {year}
  {1992})}\BibitemShut {NoStop}%
\bibitem [{\citenamefont {Perdew}\ \emph {et~al.}(1996)\citenamefont {Perdew},
  \citenamefont {Burke},\ and\ \citenamefont {Ernzerhof}}]{perdewburke}%
  \BibitemOpen
  \bibfield  {author} {\bibinfo {author} {\bibfnamefont {J.~P.}\ \bibnamefont
  {Perdew}}, \bibinfo {author} {\bibfnamefont {K.}~\bibnamefont {Burke}}, \
  and\ \bibinfo {author} {\bibfnamefont {M.}~\bibnamefont {Ernzerhof}},\
  }\bibfield  {title} {\enquote {\bibinfo {title} {Generalized gradient
  approximation made simple},}\ }\href@noop {} {\bibfield  {journal} {\bibinfo
  {journal} {Phys. Rev. Lett.}\ }\textbf {\bibinfo {volume} {77}},\ \bibinfo
  {pages} {3865--3868} (\bibinfo {year} {1996})}\BibitemShut {NoStop}%
\bibitem [{\citenamefont {Ceperly}\ and\ \citenamefont
  {Alder}(1980)}]{ceperlyadler}%
  \BibitemOpen
  \bibfield  {author} {\bibinfo {author} {\bibfnamefont {D.~M.}\ \bibnamefont
  {Ceperly}}\ and\ \bibinfo {author} {\bibfnamefont {B.~J.}\ \bibnamefont
  {Alder}},\ }\bibfield  {title} {\enquote {\bibinfo {title} {Ground state of
  the electron gas by a stochastic method},}\ }\href@noop {} {\bibfield
  {journal} {\bibinfo  {journal} {Phys. Rev. Lett.}\ }\textbf {\bibinfo
  {volume} {45}},\ \bibinfo {pages} {566--569} (\bibinfo {year}
  {1980})}\BibitemShut {NoStop}%
\bibitem [{\citenamefont {Mostofi}\ \emph {et~al.}(2014)\citenamefont
  {Mostofi}, \citenamefont {Yates}, \citenamefont {Pizzi}, \citenamefont {Lee},
  \citenamefont {Souza}, \citenamefont {Vanderbilt},\ and\ \citenamefont
  {Marzari}}]{wannier90citation1}%
  \BibitemOpen
  \bibfield  {author} {\bibinfo {author} {\bibfnamefont {A.~A.}\ \bibnamefont
  {Mostofi}}, \bibinfo {author} {\bibfnamefont {J.~R.}\ \bibnamefont {Yates}},
  \bibinfo {author} {\bibfnamefont {G.}~\bibnamefont {Pizzi}}, \bibinfo
  {author} {\bibfnamefont {Y.-S.}\ \bibnamefont {Lee}}, \bibinfo {author}
  {\bibfnamefont {I.}~\bibnamefont {Souza}}, \bibinfo {author} {\bibfnamefont
  {D.}~\bibnamefont {Vanderbilt}}, \ and\ \bibinfo {author} {\bibfnamefont
  {N.}~\bibnamefont {Marzari}},\ }\bibfield  {title} {\enquote {\bibinfo
  {title} {An updated version of wannier90: A tool for obtaining maximally
  localised {W}annier functions},}\ }\href@noop {} {\bibfield  {journal}
  {\bibinfo  {journal} {Computer Physics Communications}\ }\textbf {\bibinfo
  {volume} {185}},\ \bibinfo {pages} {2309--2310} (\bibinfo {year}
  {2014})}\BibitemShut {NoStop}%
\bibitem [{\citenamefont {Wu}\ \emph {et~al.}(2018)\citenamefont {Wu},
  \citenamefont {Zhang}, \citenamefont {Song}, \citenamefont {Troyer},\ and\
  \citenamefont {Soluyanov}}]{wanniertools}%
  \BibitemOpen
  \bibfield  {author} {\bibinfo {author} {\bibfnamefont {Q.}~\bibnamefont
  {Wu}}, \bibinfo {author} {\bibfnamefont {S.}~\bibnamefont {Zhang}}, \bibinfo
  {author} {\bibfnamefont {H.-F.}\ \bibnamefont {Song}}, \bibinfo {author}
  {\bibfnamefont {M.}~\bibnamefont {Troyer}}, \ and\ \bibinfo {author}
  {\bibfnamefont {A.~A.}\ \bibnamefont {Soluyanov}},\ }\bibfield  {title}
  {\enquote {\bibinfo {title} {Wannier{T}ools: An open-source software package
  for novel topological materials},}\ }\href@noop {} {\bibfield  {journal}
  {\bibinfo  {journal} {Computer Physics Communications}\ }\textbf {\bibinfo
  {volume} {224}},\ \bibinfo {pages} {405--416} (\bibinfo {year}
  {2018})}\BibitemShut {NoStop}%
\bibitem [{\citenamefont {Tsirkin}(2021)}]{wannierberri}%
  \BibitemOpen
  \bibfield  {author} {\bibinfo {author} {\bibfnamefont {S.}~\bibnamefont
  {Tsirkin}},\ }\bibfield  {title} {\enquote {\bibinfo {title} {High
  performance {W}annier interpolation of {B}erry curvature and related
  quantitites with {W}annier{B}erri code},}\ }\href@noop {} {\bibfield
  {journal} {\bibinfo  {journal} {npj Computational Materials}\ }\textbf
  {\bibinfo {volume} {7}},\ \bibinfo {pages} {33} (\bibinfo {year}
  {2021})}\BibitemShut {NoStop}%
\bibitem [{\citenamefont {Vo{\v c}adlo}\ \emph {et~al.}(1999)\citenamefont
  {Vo{\v c}adlo}, \citenamefont {Price},\ and\ \citenamefont {Wood}}]{vocaldo}%
  \BibitemOpen
  \bibfield  {author} {\bibinfo {author} {\bibfnamefont {L.}~\bibnamefont
  {Vo{\v c}adlo}}, \bibinfo {author} {\bibfnamefont {G.~D.}\ \bibnamefont
  {Price}}, \ and\ \bibinfo {author} {\bibfnamefont {I.~G.}\ \bibnamefont
  {Wood}},\ }\bibfield  {title} {\enquote {\bibinfo {title} {{Crystal
  structure, compressibility and possible phase transitions in
  $\varepsilon$-FeSi studied by first-principles pseudopotential
  calculations}},}\ }\href@noop {} {\bibfield  {journal} {\bibinfo  {journal}
  {Acta Crystallographica Sect. B}\ }\textbf {\bibinfo {volume} {55}},\
  \bibinfo {pages} {484--493} (\bibinfo {year} {1999})}\BibitemShut {NoStop}%
\bibitem [{\citenamefont {Klotz}\ \emph {et~al.}(2019)\citenamefont {Klotz},
  \citenamefont {G\"otze}, \citenamefont {F\"orster}, \citenamefont {Bruin},
  \citenamefont {Wosnitza}, \citenamefont {Weber}, \citenamefont {Schmidt},
  \citenamefont {Schnelle}, \citenamefont {Geibel}, \citenamefont
  {R\"o\ss{}ler},\ and\ \citenamefont {Rosner}}]{ulrichcrg}%
  \BibitemOpen
  \bibfield  {author} {\bibinfo {author} {\bibfnamefont {J.}~\bibnamefont
  {Klotz}}, \bibinfo {author} {\bibfnamefont {K.}~\bibnamefont {G\"otze}},
  \bibinfo {author} {\bibfnamefont {T.}~\bibnamefont {F\"orster}}, \bibinfo
  {author} {\bibfnamefont {J.~A.~N.}\ \bibnamefont {Bruin}}, \bibinfo {author}
  {\bibfnamefont {J.}~\bibnamefont {Wosnitza}}, \bibinfo {author}
  {\bibfnamefont {K.}~\bibnamefont {Weber}}, \bibinfo {author} {\bibfnamefont
  {M.}~\bibnamefont {Schmidt}}, \bibinfo {author} {\bibfnamefont
  {W.}~\bibnamefont {Schnelle}}, \bibinfo {author} {\bibfnamefont
  {C.}~\bibnamefont {Geibel}}, \bibinfo {author} {\bibfnamefont {U.~K.}\
  \bibnamefont {R\"o\ss{}ler}}, \ and\ \bibinfo {author} {\bibfnamefont
  {H.}~\bibnamefont {Rosner}},\ }\bibfield  {title} {\enquote {\bibinfo {title}
  {Electronic band structure and proximity to magnetic ordering in the chiral
  cubic compound {C}r{G}e},}\ }\href@noop {} {\bibfield  {journal} {\bibinfo
  {journal} {Phys. Rev. B}\ }\textbf {\bibinfo {volume} {99}},\ \bibinfo
  {pages} {085130} (\bibinfo {year} {2019})}\BibitemShut {NoStop}%
\bibitem [{\citenamefont {Bradlyn}\ \emph {et~al.}(2016)\citenamefont
  {Bradlyn}, \citenamefont {Cano}, \citenamefont {Wang}, \citenamefont
  {Vergniory}, \citenamefont {Felser}, \citenamefont {Cava},\ and\
  \citenamefont {Bernevig}}]{bradlyn}%
  \BibitemOpen
  \bibfield  {author} {\bibinfo {author} {\bibfnamefont {B.}~\bibnamefont
  {Bradlyn}}, \bibinfo {author} {\bibfnamefont {J.}~\bibnamefont {Cano}},
  \bibinfo {author} {\bibfnamefont {Z.}~\bibnamefont {Wang}}, \bibinfo {author}
  {\bibfnamefont {M.~G.}\ \bibnamefont {Vergniory}}, \bibinfo {author}
  {\bibfnamefont {C.}~\bibnamefont {Felser}}, \bibinfo {author} {\bibfnamefont
  {R.~J.}\ \bibnamefont {Cava}}, \ and\ \bibinfo {author} {\bibfnamefont
  {B.~A.}\ \bibnamefont {Bernevig}},\ }\bibfield  {title} {\enquote {\bibinfo
  {title} {Beyond {D}irac and {W}eyl fermions: Unconventional quasiparticles in
  conventional crystals},}\ }\href@noop {} {\bibfield  {journal} {\bibinfo
  {journal} {Science}\ }\textbf {\bibinfo {volume} {353}},\ \bibinfo {pages}
  {aaf5037} (\bibinfo {year} {2016})}\BibitemShut {NoStop}%
\bibitem [{\citenamefont {Huber}\ \emph {et~al.}(2022)\citenamefont {Huber},
  \citenamefont {Alpin}, \citenamefont {Causer}, \citenamefont {Worch},
  \citenamefont {Bauer}, \citenamefont {Benka}, \citenamefont {Hirschmann},
  \citenamefont {Schnyder}, \citenamefont {Pfleiderer},\ and\ \citenamefont
  {Wilde}}]{nico}%
  \BibitemOpen
  \bibfield  {author} {\bibinfo {author} {\bibfnamefont {N.}~\bibnamefont
  {Huber}}, \bibinfo {author} {\bibfnamefont {K.}~\bibnamefont {Alpin}},
  \bibinfo {author} {\bibfnamefont {G.~L.}\ \bibnamefont {Causer}}, \bibinfo
  {author} {\bibfnamefont {L.}~\bibnamefont {Worch}}, \bibinfo {author}
  {\bibfnamefont {A.}~\bibnamefont {Bauer}}, \bibinfo {author} {\bibfnamefont
  {G.}~\bibnamefont {Benka}}, \bibinfo {author} {\bibfnamefont {M.~M.}\
  \bibnamefont {Hirschmann}}, \bibinfo {author} {\bibfnamefont {A.~P.}\
  \bibnamefont {Schnyder}}, \bibinfo {author} {\bibfnamefont {C.}~\bibnamefont
  {Pfleiderer}}, \ and\ \bibinfo {author} {\bibfnamefont {M.~A.}\ \bibnamefont
  {Wilde}},\ }\bibfield  {title} {\enquote {\bibinfo {title} {{Network of
  Topological Nodal Planes, Multifold Degeneracies, and Weyl Points in
  CoSi}},}\ }\href@noop {} {\bibfield  {journal} {\bibinfo  {journal} {Phys.
  Rev. Lett.}\ }\textbf {\bibinfo {volume} {129}},\ \bibinfo {pages} {026401}
  (\bibinfo {year} {2022})}\BibitemShut {NoStop}%
\bibitem [{\citenamefont {Tran}\ \emph {et~al.}(2007)\citenamefont {Tran},
  \citenamefont {Laskowski}, \citenamefont {Blaha},\ and\ \citenamefont
  {Schwarz}}]{wucu}%
  \BibitemOpen
  \bibfield  {author} {\bibinfo {author} {\bibfnamefont {F.}~\bibnamefont
  {Tran}}, \bibinfo {author} {\bibfnamefont {R.}~\bibnamefont {Laskowski}},
  \bibinfo {author} {\bibfnamefont {P.}~\bibnamefont {Blaha}}, \ and\ \bibinfo
  {author} {\bibfnamefont {K.}~\bibnamefont {Schwarz}},\ }\bibfield  {title}
  {\enquote {\bibinfo {title} {{Performance on molecules, surfaces, and solids
  of the Wu-Cohen GGA exchange-correlation energy functional}},}\ }\href@noop
  {} {\bibfield  {journal} {\bibinfo  {journal} {Phys. Rev. B}\ }\textbf
  {\bibinfo {volume} {75}},\ \bibinfo {pages} {115131} (\bibinfo {year}
  {2007})}\BibitemShut {NoStop}%
\bibitem [{\citenamefont {Grabowski}\ \emph {et~al.}(2007)\citenamefont
  {Grabowski}, \citenamefont {Hickel},\ and\ \citenamefont
  {Neugebauer}}]{ldaunderestimation}%
  \BibitemOpen
  \bibfield  {author} {\bibinfo {author} {\bibfnamefont {B.}~\bibnamefont
  {Grabowski}}, \bibinfo {author} {\bibfnamefont {T.}~\bibnamefont {Hickel}}, \
  and\ \bibinfo {author} {\bibfnamefont {J.}~\bibnamefont {Neugebauer}},\
  }\bibfield  {title} {\enquote {\bibinfo {title} {Ab initio study of the
  thermodynamic properties of nonmagnetic elementary fcc metals:
  Exchange-correlation-related error bars and chemical trends},}\ }\href@noop
  {} {\bibfield  {journal} {\bibinfo  {journal} {Phys. Rev. B}\ }\textbf
  {\bibinfo {volume} {76}},\ \bibinfo {pages} {024309} (\bibinfo {year}
  {2007})}\BibitemShut {NoStop}%
\bibitem [{\citenamefont {Hebbache}\ and\ \citenamefont
  {Zemzemi}(2004)}]{vinetequationofstate}%
  \BibitemOpen
  \bibfield  {author} {\bibinfo {author} {\bibfnamefont {M.}~\bibnamefont
  {Hebbache}}\ and\ \bibinfo {author} {\bibfnamefont {M.}~\bibnamefont
  {Zemzemi}},\ }\bibfield  {title} {\enquote {\bibinfo {title} {{Ab initio
  study of high-pressure behaviour of a low compressibility metal and a hard
  material: Osmium and diamond}},}\ }\href@noop {} {\bibfield  {journal}
  {\bibinfo  {journal} {Phys. Rev. B}\ }\textbf {\bibinfo {volume} {70}},\
  \bibinfo {pages} {224107} (\bibinfo {year} {2004})}\BibitemShut {NoStop}%
\bibitem [{\citenamefont {Choudhary}\ \emph {et~al.}(2018)\citenamefont
  {Choudhary}, \citenamefont {Cheon}, \citenamefont {Reed},\ and\ \citenamefont
  {Tavazza}}]{bulkmodulus}%
  \BibitemOpen
  \bibfield  {author} {\bibinfo {author} {\bibfnamefont {K.}~\bibnamefont
  {Choudhary}}, \bibinfo {author} {\bibfnamefont {G.}~\bibnamefont {Cheon}},
  \bibinfo {author} {\bibfnamefont {E.}~\bibnamefont {Reed}}, \ and\ \bibinfo
  {author} {\bibfnamefont {F.}~\bibnamefont {Tavazza}},\ }\bibfield  {title}
  {\enquote {\bibinfo {title} {{Elastic properties of bulk and low-dimensional
  materials using van der Waals density functional}},}\ }\href@noop {}
  {\bibfield  {journal} {\bibinfo  {journal} {Phys. Rev. B}\ }\textbf {\bibinfo
  {volume} {98}},\ \bibinfo {pages} {014107} (\bibinfo {year}
  {2018})}\BibitemShut {NoStop}%
\bibitem [{\citenamefont {Bai}\ \emph {et~al.}(2020)\citenamefont {Bai},
  \citenamefont {Li}, \citenamefont {Xiao}, \citenamefont {Rao}, \citenamefont
  {Liang}, \citenamefont {He},\ and\ \citenamefont {Feng}}]{bulkmodulusref}%
  \BibitemOpen
  \bibfield  {author} {\bibinfo {author} {\bibfnamefont {X.}~\bibnamefont
  {Bai}}, \bibinfo {author} {\bibfnamefont {Y.}~\bibnamefont {Li}}, \bibinfo
  {author} {\bibfnamefont {B.}~\bibnamefont {Xiao}}, \bibinfo {author}
  {\bibfnamefont {Y.}~\bibnamefont {Rao}}, \bibinfo {author} {\bibfnamefont
  {H.}~\bibnamefont {Liang}}, \bibinfo {author} {\bibfnamefont
  {L.}~\bibnamefont {He}}, \ and\ \bibinfo {author} {\bibfnamefont
  {J.}~\bibnamefont {Feng}},\ }\bibfield  {title} {\enquote {\bibinfo {title}
  {{Structural, mechanical, electronic properties of refractory Hf-Al
  intermetallics from SCAN meta-GGA density functional calculations}},}\
  }\href@noop {} {\bibfield  {journal} {\bibinfo  {journal} {Materials
  Chemistry and Physics}\ }\textbf {\bibinfo {volume} {254}},\ \bibinfo {pages}
  {123423} (\bibinfo {year} {2020})}\BibitemShut {NoStop}%
\bibitem [{\citenamefont {Fu}\ and\ \citenamefont
  {Ho}(1983)}]{murnaghanequationofstate}%
  \BibitemOpen
  \bibfield  {author} {\bibinfo {author} {\bibfnamefont {C.-L.}\ \bibnamefont
  {Fu}}\ and\ \bibinfo {author} {\bibfnamefont {K.-M.}\ \bibnamefont {Ho}},\
  }\bibfield  {title} {\enquote {\bibinfo {title} {{First-principles
  calculation of the equlibrium ground-state properties of transition metals:
  Applications to Nb and Mo}},}\ }\href@noop {} {\bibfield  {journal} {\bibinfo
   {journal} {Phys. Rev. B}\ }\textbf {\bibinfo {volume} {28}},\ \bibinfo
  {pages} {5480--5486} (\bibinfo {year} {1983})}\BibitemShut {NoStop}%
\bibitem [{\citenamefont {Murnaghan}(1994)}]{inversebmequation}%
  \BibitemOpen
  \bibfield  {author} {\bibinfo {author} {\bibfnamefont {F.~D.}\ \bibnamefont
  {Murnaghan}},\ }\bibfield  {title} {\enquote {\bibinfo {title} {The
  compressibility of media under extreme pressures},}\ }\href@noop {}
  {\bibfield  {journal} {\bibinfo  {journal} {Proceedings of the National
  Academy of Sciences}\ }\textbf {\bibinfo {volume} {30}},\ \bibinfo {pages}
  {244--247} (\bibinfo {year} {1994})}\BibitemShut {NoStop}%
\bibitem [{\citenamefont {Barber}(2018)}]{barber}%
  \BibitemOpen
  \bibfield  {author} {\bibinfo {author} {\bibfnamefont {M.}~\bibnamefont
  {Barber}},\ }\href@noop {} {\emph {\bibinfo {title} {Uniaxial Stress
  Technique and Investigations of Correlated Electron Systems}}},\ \bibinfo
  {number} {978-3-319-93972-8}\ (\bibinfo  {publisher} {Springer International
  Publishing},\ \bibinfo {year} {2018})\ pp.\ \bibinfo {pages}
  {13--48}\BibitemShut {NoStop}%
\bibitem [{\citenamefont {Ziman}(1979)}]{vcabasic}%
  \BibitemOpen
  \bibfield  {author} {\bibinfo {author} {\bibfnamefont {J.~M.}\ \bibnamefont
  {Ziman}},\ }\href@noop {} {\emph {\bibinfo {title} {Models of Disorder: The
  Theoretical Physics of Homogeneously Disordered Systems}}},\ \bibinfo
  {number} {978-0521292801}\ (\bibinfo  {publisher} {Cambridge University
  Press},\ \bibinfo {year} {1979})\BibitemShut {NoStop}%
\bibitem [{\citenamefont {Barman}\ \emph {et~al.}(2020)\citenamefont {Barman},
  \citenamefont {Mondal}, \citenamefont {Pujari}, \citenamefont {Pathak},\ and\
  \citenamefont {Alam}}]{coge}%
  \BibitemOpen
  \bibfield  {author} {\bibinfo {author} {\bibfnamefont {C.~K.}\ \bibnamefont
  {Barman}}, \bibinfo {author} {\bibfnamefont {C.}~\bibnamefont {Mondal}},
  \bibinfo {author} {\bibfnamefont {S.}~\bibnamefont {Pujari}}, \bibinfo
  {author} {\bibfnamefont {B.}~\bibnamefont {Pathak}}, \ and\ \bibinfo {author}
  {\bibfnamefont {A.}~\bibnamefont {Alam}},\ }\bibfield  {title} {\enquote
  {\bibinfo {title} {{Symmetry protection and giant Fermi arcs from multifold
  fermions in binary, ternary, and quaternary compounds}},}\ }\href@noop {}
  {\bibfield  {journal} {\bibinfo  {journal} {Phys. Rev. B}\ }\textbf {\bibinfo
  {volume} {102}},\ \bibinfo {pages} {155147} (\bibinfo {year}
  {2020})}\BibitemShut {NoStop}%
\bibitem [{\citenamefont {Pshenay-Severin}\ \emph {et~al.}(2018)\citenamefont
  {Pshenay-Severin}, \citenamefont {Ivanov}, \citenamefont {Burkov},\ and\
  \citenamefont {Burkov}}]{cosiseverin}%
  \BibitemOpen
  \bibfield  {author} {\bibinfo {author} {\bibfnamefont {D.~A.}\ \bibnamefont
  {Pshenay-Severin}}, \bibinfo {author} {\bibfnamefont {Y.~V.}\ \bibnamefont
  {Ivanov}}, \bibinfo {author} {\bibfnamefont {A.~A.}\ \bibnamefont {Burkov}},
  \ and\ \bibinfo {author} {\bibfnamefont {A.~T.}\ \bibnamefont {Burkov}},\
  }\bibfield  {title} {\enquote {\bibinfo {title} {{Band structure and
  unconventional electronic topology of CoSi}},}\ }\href@noop {} {\bibfield
  {journal} {\bibinfo  {journal} {Journal of Physics: Condensed Matter}\
  }\textbf {\bibinfo {volume} {30}},\ \bibinfo {pages} {135501} (\bibinfo
  {year} {2018})}\BibitemShut {NoStop}%
\bibitem [{\citenamefont {Yang}\ \emph {et~al.}(2023)\citenamefont {Yang},
  \citenamefont {Xiao}, \citenamefont {Robredo}, \citenamefont {Vergniory},
  \citenamefont {Yan},\ and\ \citenamefont {Felser}}]{orbitaltexture}%
  \BibitemOpen
  \bibfield  {author} {\bibinfo {author} {\bibfnamefont {Q.}~\bibnamefont
  {Yang}}, \bibinfo {author} {\bibfnamefont {J.}~\bibnamefont {Xiao}}, \bibinfo
  {author} {\bibfnamefont {I.}~\bibnamefont {Robredo}}, \bibinfo {author}
  {\bibfnamefont {M.~G.}\ \bibnamefont {Vergniory}}, \bibinfo {author}
  {\bibfnamefont {B.}~\bibnamefont {Yan}}, \ and\ \bibinfo {author}
  {\bibfnamefont {C.}~\bibnamefont {Felser}},\ }\bibfield  {title} {\enquote
  {\bibinfo {title} {Monopol-like orbital momentum locking and the induced
  orbital transport in topological chiral semimetals},}\ }\href@noop {}
  {\bibfield  {journal} {\bibinfo  {journal} {Proceedings of the National
  Academy of Sciences}\ }\textbf {\bibinfo {volume} {120}},\ \bibinfo {pages}
  {e2305541120} (\bibinfo {year} {2023})}\BibitemShut {NoStop}%
\bibitem [{\citenamefont {Kaushik}\ and\ \citenamefont
  {Cano}(2021)}]{opticalresponse}%
  \BibitemOpen
  \bibfield  {author} {\bibinfo {author} {\bibfnamefont {S.}~\bibnamefont
  {Kaushik}}\ and\ \bibinfo {author} {\bibfnamefont {J.}~\bibnamefont {Cano}},\
  }\bibfield  {title} {\enquote {\bibinfo {title} {{Magnetic photocurrents in
  multifold Weyl fermions}},}\ }\href@noop {} {\bibfield  {journal} {\bibinfo
  {journal} {Phys. Rev. B}\ }\textbf {\bibinfo {volume} {104}},\ \bibinfo
  {pages} {155149} (\bibinfo {year} {2021})}\BibitemShut {NoStop}%
\bibitem [{\citenamefont {de~Juan}\ \emph {et~al.}(2017)\citenamefont
  {de~Juan}, \citenamefont {Grushin}, \citenamefont {Morimoto},\ and\
  \citenamefont {Moore}}]{quantized}%
  \BibitemOpen
  \bibfield  {author} {\bibinfo {author} {\bibfnamefont {F.}~\bibnamefont
  {de~Juan}}, \bibinfo {author} {\bibfnamefont {A.~G.}\ \bibnamefont
  {Grushin}}, \bibinfo {author} {\bibfnamefont {T.}~\bibnamefont {Morimoto}}, \
  and\ \bibinfo {author} {\bibfnamefont {J.~E.}\ \bibnamefont {Moore}},\
  }\bibfield  {title} {\enquote {\bibinfo {title} {{Quantized circular
  photogalvanic effect In Weyl semimetals}},}\ }\href@noop {} {\bibfield
  {journal} {\bibinfo  {journal} {Nature Communications}\ }\textbf {\bibinfo
  {volume} {8}},\ \bibinfo {pages} {15995} (\bibinfo {year}
  {2017})}\BibitemShut {NoStop}%
\bibitem [{\citenamefont {Nielsen}\ and\ \citenamefont
  {Ninomiya}(1983)}]{anomaly}%
  \BibitemOpen
  \bibfield  {author} {\bibinfo {author} {\bibfnamefont {H.~B.}\ \bibnamefont
  {Nielsen}}\ and\ \bibinfo {author} {\bibfnamefont {M.}~\bibnamefont
  {Ninomiya}},\ }\bibfield  {title} {\enquote {\bibinfo {title} {{The
  Adler-Bell-Jackiw anomaly and Weyl fermions in a crystal}},}\ }\href@noop {}
  {\bibfield  {journal} {\bibinfo  {journal} {Physics Letters B}\ }\textbf
  {\bibinfo {volume} {130}},\ \bibinfo {pages} {389--396} (\bibinfo {year}
  {1983})}\BibitemShut {NoStop}%
\end{thebibliography}
%

\newpage
\section{Appendix}
\subsection{Wannier fit}
\label{Wannierfit}
\begin{figure}[H]
\centering
\includegraphics[width=0.52\textwidth, keepaspectratio]{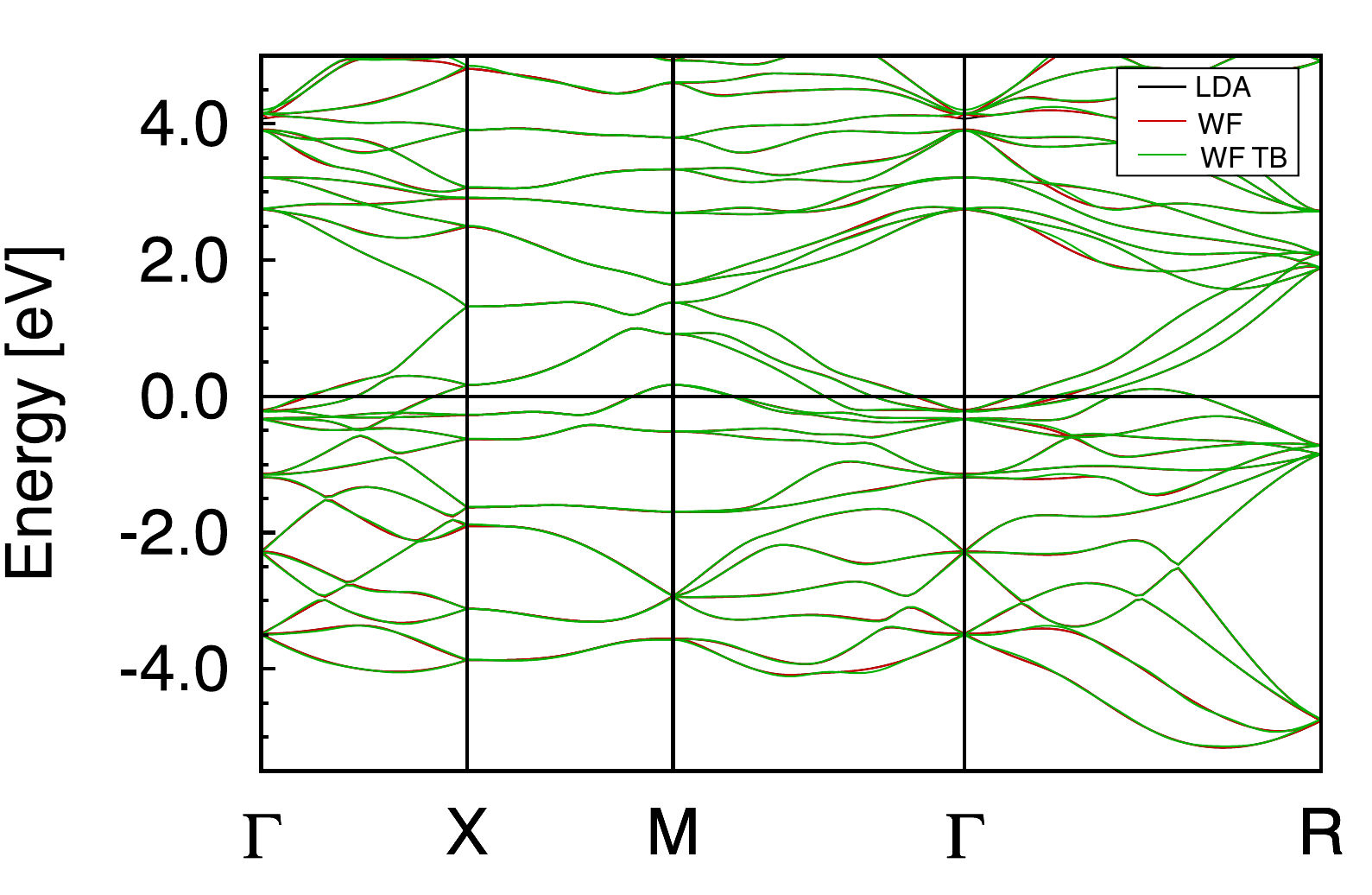}
\caption{Comparison of full DFT band structure with the one from reduced Wannier model when SOC is not included.}
\label{wannierfitwithoutSOC}
\end{figure}

\begin{figure}[H]
\centering
\includegraphics[width=0.52\textwidth, keepaspectratio]{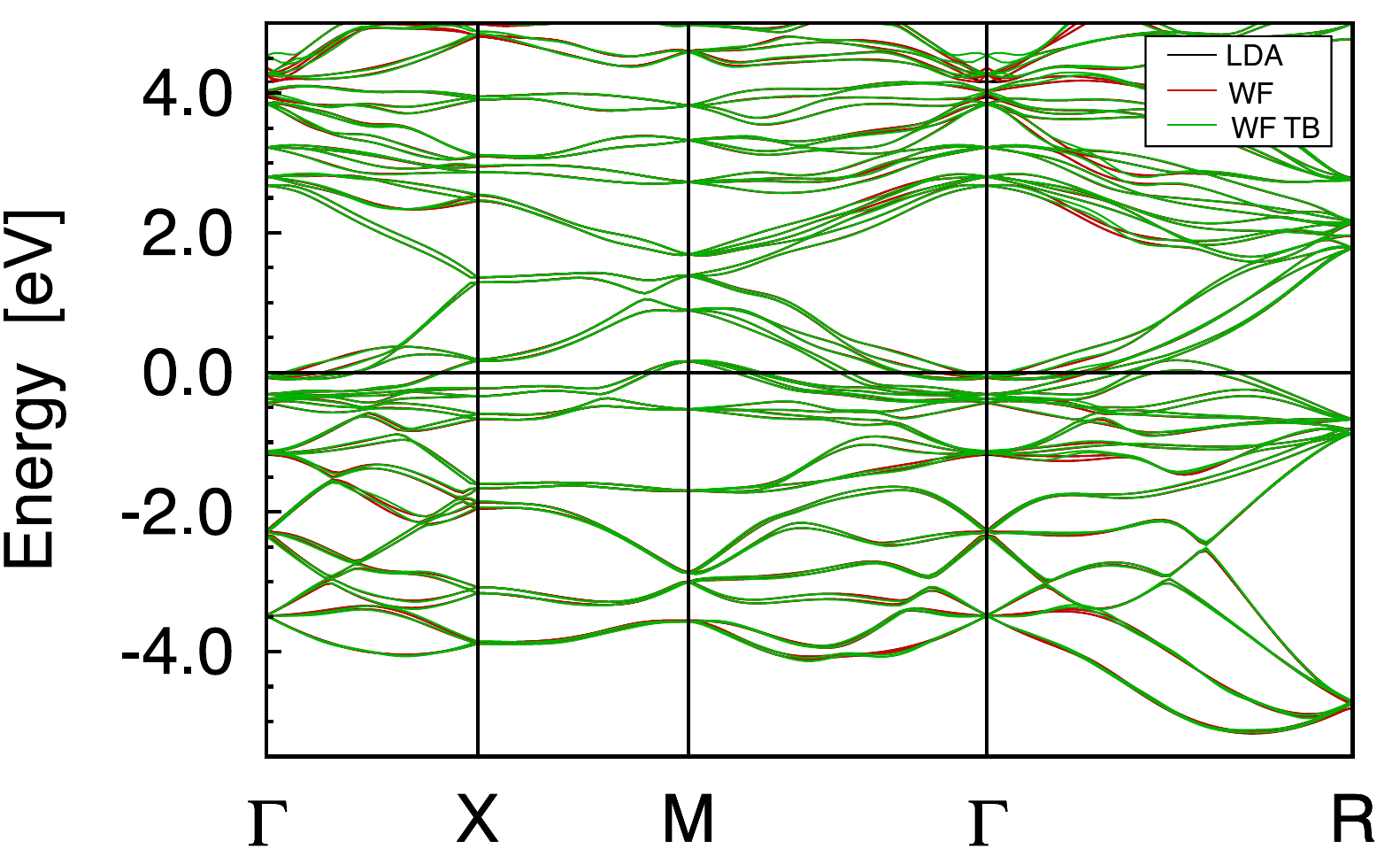}
\caption{Comparison of DFT band structure and Wannier fit with SOC.}
\label{wannierfitwithSOC}
\end{figure}

\hspace*{2em} Figures $\ref{wannierfitwithoutSOC}$ and \ref{wannierfitwithSOC} show the comparison of the DFT LDA band structure (black) with the one from the Wannier function model (red). The green curve represents the TB approximation to the full WF model (red), where cutoff for the hoppings and magnitude threshold are taken into account. It is visible that the regions around the Fermi level are well reproduced (red and black curves are on top of each other in both figures in a given energy range). There exist minor deviations of the WF TB model from the DFT band structure. However, these are not of any relevance for the results presented in the current study. Good qualitative agreement of the DFT, WF and WF TB band structures is a strong evidence that all relevant hopping parameters are properly taken into account.\\
\subsection{LDA versus GGA}
\label{LDAversusGGAcomparison}
\begin{figure}[H]
\centering
\includegraphics[width=0.52\textwidth, keepaspectratio]{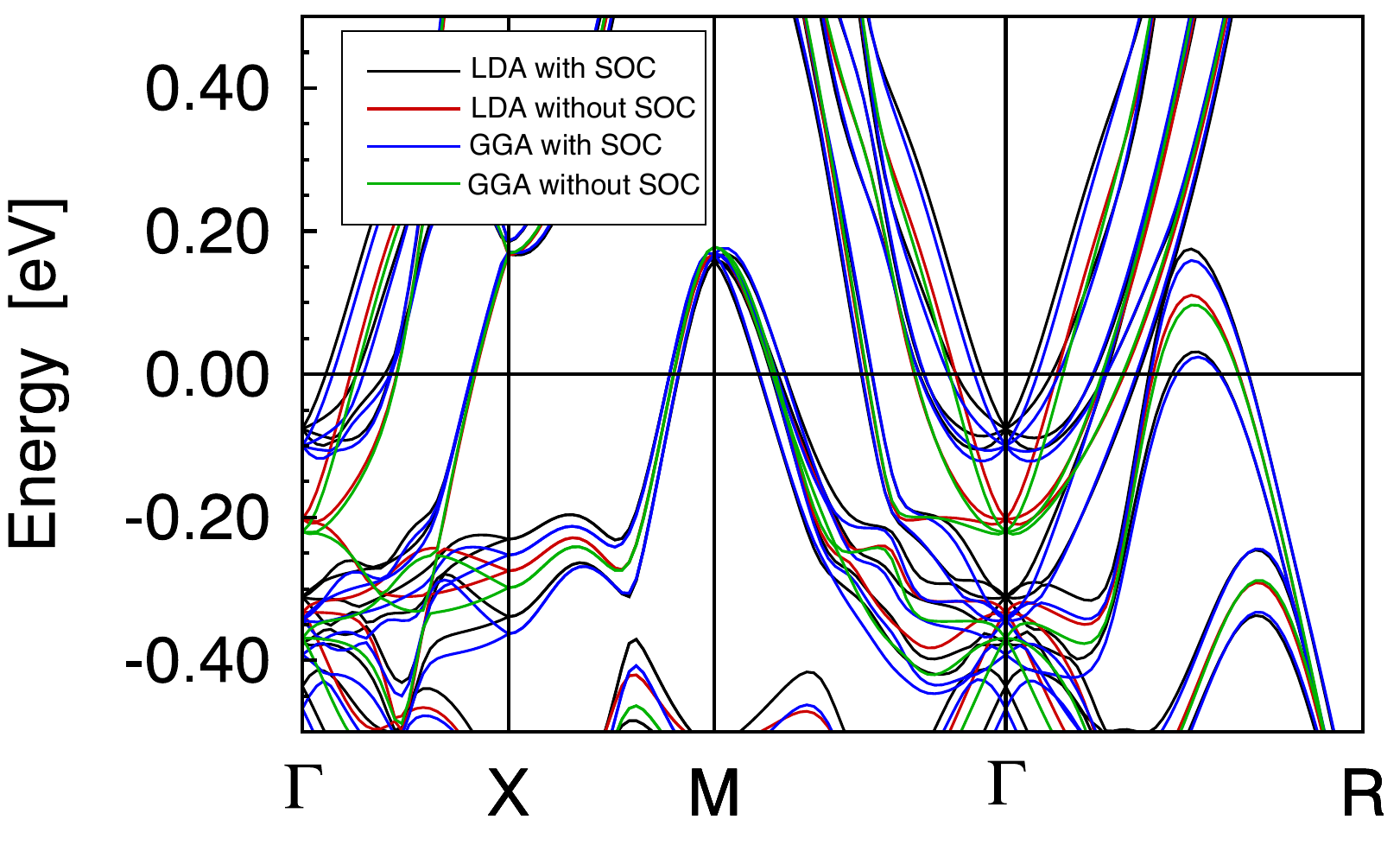}
\caption{Comparison of LDA and GGA band structures with and without SOC.}
\label{LDAGGAwithandwithoutSOC}
\end{figure}

\hspace*{2em} In Fi. $\ref{LDAGGAwithandwithoutSOC}$ we show a comparison of the band structures from LDA and GGA with and without SOC. While both functionals provide the same general description, the GGA eigenvalues are shifted down compared to the LDA values at our point of interest ($\Gamma$). The threefold degeneracy at $\Gamma$ in the calculations with neglected SOC becomes fourfold and twofold degenerate with SOC, and it is lifted up in energy in both functionals.
\subsection{Optimization of the exp. structure}
\begin{figure}[H]
\centering
\includegraphics[width=0.52\textwidth, keepaspectratio]{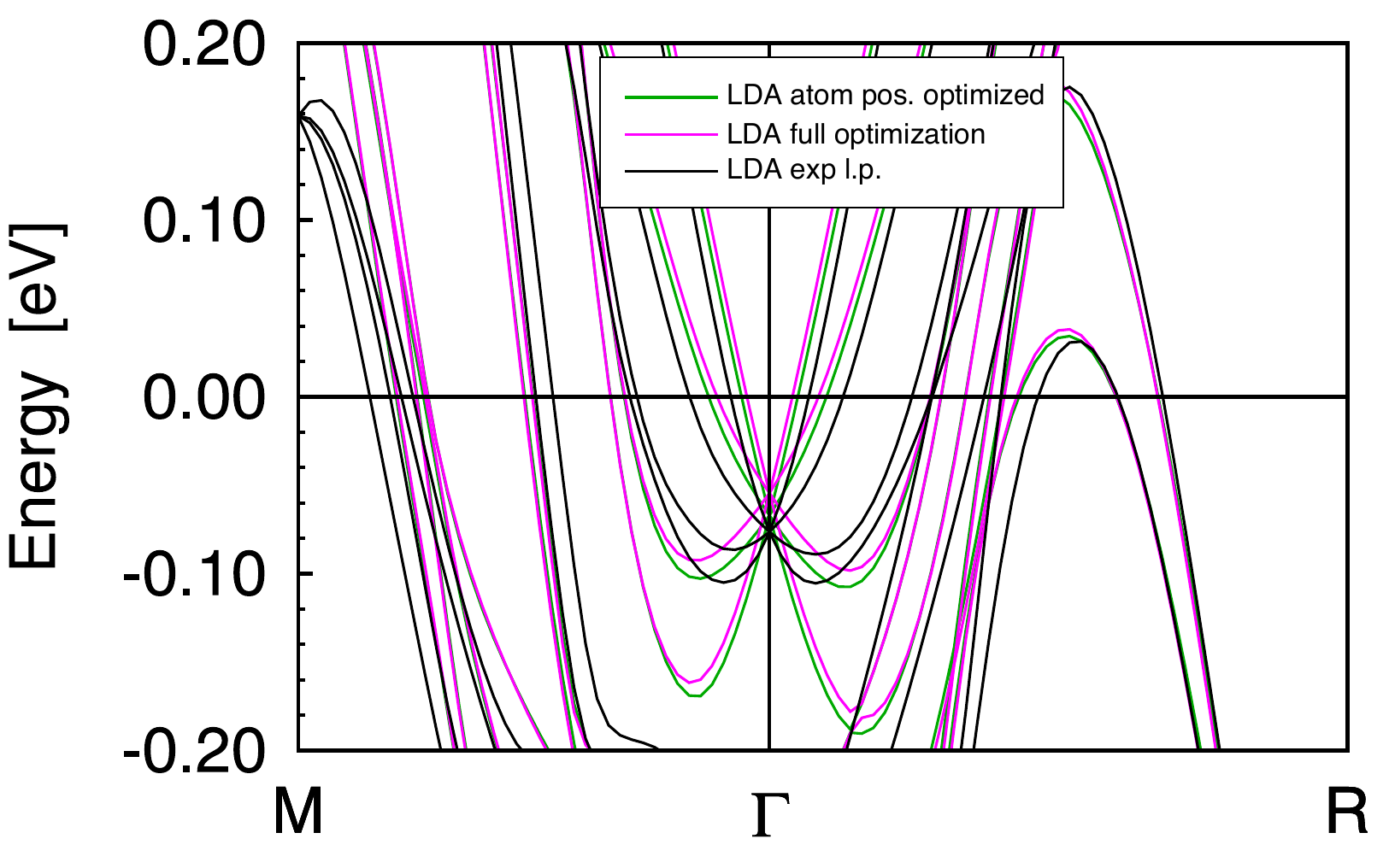}
\caption{Zoom around the E$_{F}$ of LDA bands with SOC for exp., atom positions optimized and fully optimized structure.}
\label{LDAfullandatomopt}
\end{figure}

In Fig. $\ref{LDAfullandatomopt}$ we compare the band structures obtained by two steps of crystal structure optimization (full and atomic coordinates only) with the experimental crystal structure.
\subsection{Volume-pressure dependence}
\begin{figure}[h!]
\centering
\includegraphics[width=0.52\textwidth, keepaspectratio]{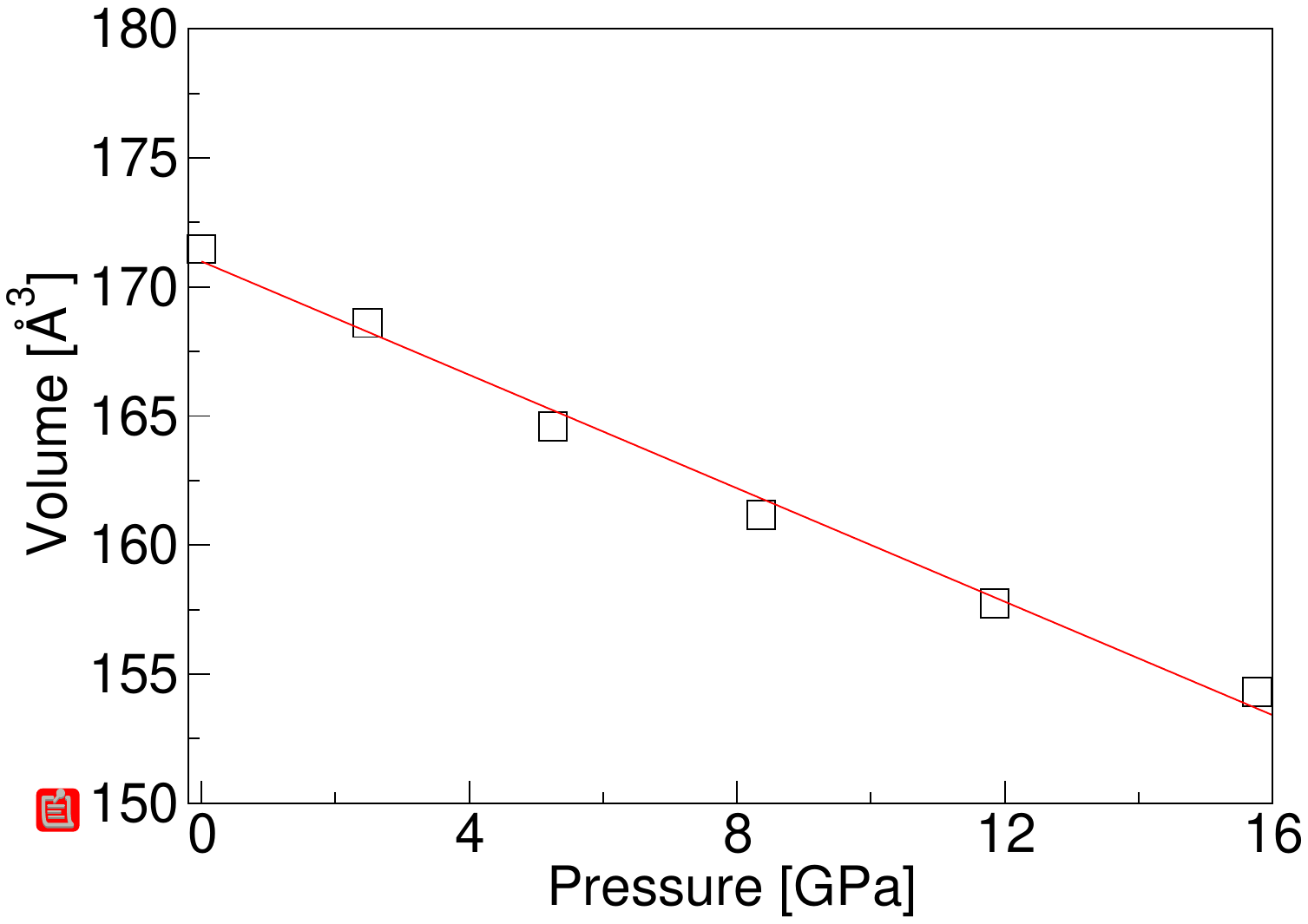}
\caption{Pressure dependence of the unit cell volume.}
\label{volumeversuspressuredependence}
\end{figure}

Fig. $\ref{volumeversuspressuredependence}$ shows the calculated volume-pressure dependence for HfSn, obtained by making use of inverse Birch-Murnaghan equation \cite{inversebmequation}.
\subsection{Berry curvature}
\label{berrycurvature}
\begin{figure}[H]
\centering
\includegraphics[width=0.45\textwidth, keepaspectratio]{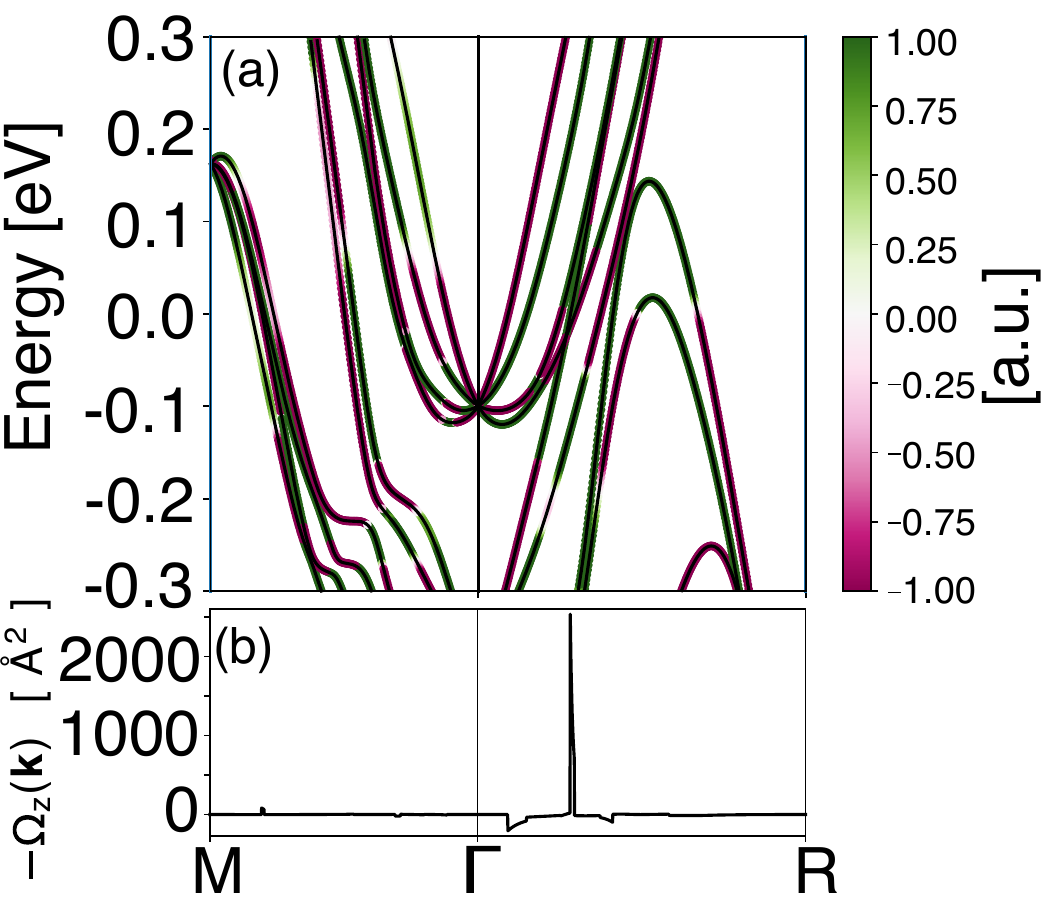}
\caption{(a) Band resolved BC along M-$\Gamma$-R; (b) The BC summed over all occupied states below the Fermi energy for the same path.}
\label{berry}
\end{figure}

Fig. $\ref{berry}$ shows the band-resolved and the Berry curvature (BC) summed over the occupied states. The color coding is according to the z-component of the BC. We observe a large peak in the BC (Fig. \ref{berry} (b)) originating from two almost degenerate bands at the E$_{F}$ along $\Gamma$-R direction. In addition, along the same direction there are two smaller peaks originating from the bands in the vicinity of the Fermi level, created as a result of SOC.\\
\begin{figure}[H]
\centering
\includegraphics[width=0.45\textwidth, keepaspectratio]{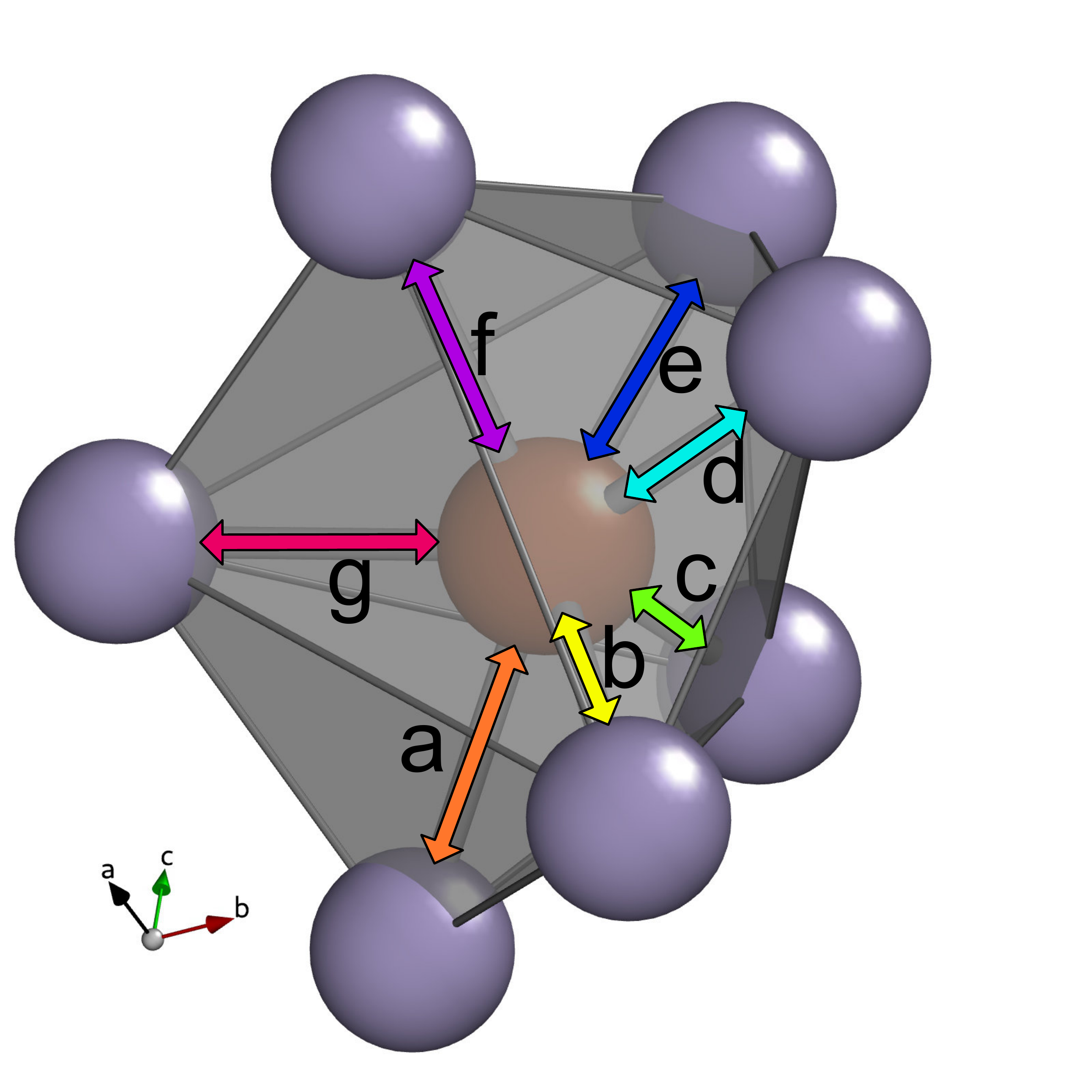}
\caption{NN polyhedron of the HfSn. The letters denote different hopping directions that can occur in the NN polyhedron.}
\label{NNpolyhedronwithlabels}
\end{figure}
\subsection{Parameters of an effective tight-binding model}
\label{tableswithandwithoutsoc}
\begin{figure}[H]
\centering
\includegraphics[width=0.5\textwidth, keepaspectratio]{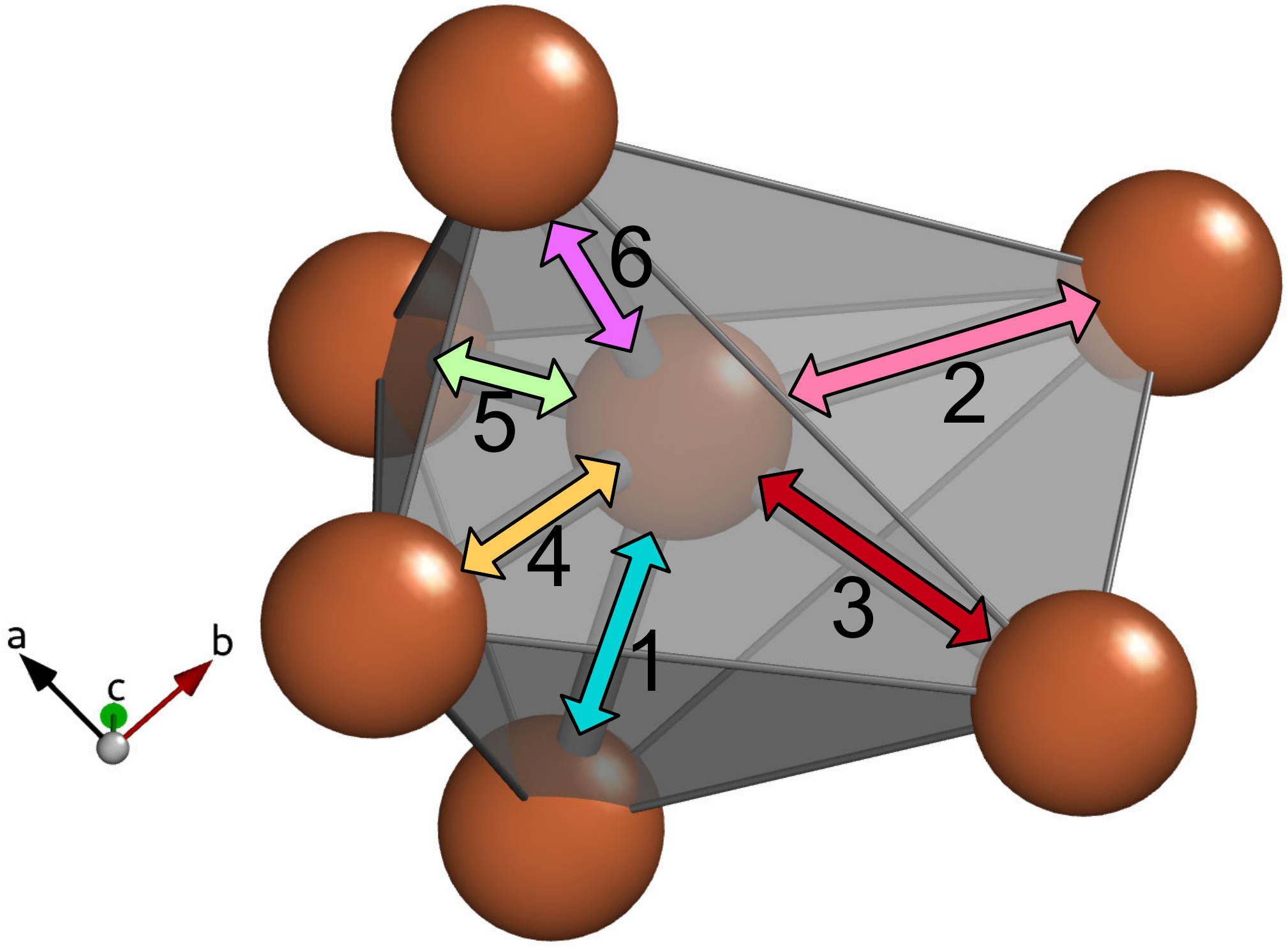}
\caption{NNN polyhedron of the HfSn. There are six possibilites for the hopping from one Hf to the neighbouring ones denoted with numbers from 1 to 6.}
\label{NNNpolyhedronwithlabels}
\end{figure}

\hspace*{2em} In Tables below we provide hopping parameters for calculations with and without SOC. The superscript indices stand for different hopping directions and notation is according to Fig. $\ref{NNpolyhedronwithlabels}$ for nearest neighbours and Fig. $\ref{NNNpolyhedronwithlabels}$ for next nearest neighbours. Due to the fact that HfSn is not expected to show heavy relativistic effects, the full relativistic basis in calculations with SOC can be projected onto a $lm\sigma$ basis in a good approximation. To distinguish orbital projections in calculations without SOC and with SOC, for the latter case we employ $\sim$ symbol above the orbitals.
\begin{table*}[h!]
\begin{ruledtabular}
\begin{tabular}{|c|c|c|c|c|c|c|c|c|}\toprule
    & Hf $5$d$_{xy}$ & Hf $5$d$_{yz}$ & Hf $5$d$_{z^{2}}$& Hf $5$d$_{xz}$& Hf $5$d$_{x^{2}-y^{2}}$&Hf $6$p$_{y}$ &Hf $6$p$_{z}$ &Hf $6$p$_{x}$\\
    \hline
 Sn  $5$p$_{y}$   &         \makecell{${-1.17}^g$\\${-1.11}^e$\\${-0.49}^{c}$\\${0.46}^{f}$\\${0.42}^{a}$}                &       \makecell{$0.48^{a}$\\${0.26}^{f}$}                     &            \makecell{${0.87}^{d}$\\${-0.86}^{g}$\\${-0.77}^{e}$\\${-0.75}^{b}$}                &          ${0.81}^{a}$                &      \makecell{${-0.95}^{g}$\\${0.91}^{d}$}                    &      \makecell{${-1.70}^{g}$\\${-1.47}^{d}$\\${-0.55}^{a}$\\${0.36}^{c}$}              &       \makecell{${-1.01}^{e}$\\${-0.74}^{a}$\\${0.71}^{b}$\\${-0.37}^{c}$}       & \makecell{${-0.78}^{d}$\\${-0.65}^{a}$\\${0.60}^{g}$\\${-0.36}^{b}$}                     \\
 \hline
 Sn  $5$p$_{z}$   &          ${0.81}^{a}$                &     \makecell{${-1.17}^{b}$\\${-1.11}^{e}$\\${-0.49}^{g}$\\${0.46}^{d}$\\${0.42}^{a}$}                       &          \makecell{${1.25}^{b}$\\${-1.22}^{e}$\\${0.48}^{c}$\\${0.39}^{f}$}                  &      \makecell{${0.48}^{a}$\\${0.26}^{d}$}                    &                 \makecell{${-0.67}^{f}$\\${-0.59}^{c}$\\${-0.27}^{b}$\\${0.29}^{e}$}         &   \makecell{${-0.78}^{e}$\\${-0.65}^{a}$\\${0.60}^{b}$\\${-0.36}^{c}$}                    &      \makecell{${-1.70}^{b}$\\${-1.47}^{e}$\\${-0.55}^{a}$\\${0.36}^{g}$}                  &     \makecell{${-1.01}^{f}$\\${-0.74}^{a}$\\${0.71}^{c}$\\${-0.37}^{g}$}                  \\
 \hline
 Sn  $5$p$_{x}$   &          \makecell{${0.48}^{a}$\\${0.26}^{e}$}             &               ${0.81}^{a}$            &           \makecell{${-0.39}^{c}$\\${0.38}^{d}$\\${0.36}^{f}$\\${0.27}^{g}$}                  &           \makecell{${-1.17}^{c}$\\${-1.11}^{f}$\\${-0.49}^{b}$\\${0.46}^{e}$\\${0.42}^{a}$}               &       \makecell{${1.22}^{c}$\\${-1.21}^{f}$\\${0.71}^{g}$\\${0.67}^{d}$}                   &    \makecell{${-1.01}^{d}$\\${-0.74}^{a}$\\${0.71}^{g}$\\${-0.37}^{b}$}                    &     \makecell{${-0.78}^{f}$\\${-0.65}^{a}$\\${0.60}^{c}$\\${-0.36}^{g}$}                  &                         \makecell{${-1.70}^{c}$\\${-1.47}^{f}$\\${-0.55}^{a}$\\${0.36}^{b}$}
\end{tabular}
\end{ruledtabular}
\caption{NN hopping parameters in eV for LDA calculations without SOC. The hopping parameters smaller than $0.25$ eV are not presented in the table.}
\label{tableofnearestneighbourhoppingparameters}
\end{table*}

\begin{table*}[h!]
\begin{ruledtabular}
\begin{tabular}{|c|c|c|c|c|c|c|c|c|}\toprule
    & Hf $5$d$_{xy}$ & Hf $5$d$_{yz}$ & Hf $5$d$_{z^{2}}$& Hf $5$d$_{xz}$& Hf $5$d$_{x^{2}-y^{2}}$&Hf $6$p$_{y}$ &Hf $6$p$_{z}$ &Hf $6$p$_{x}$\\
    \hline
 Hf  $5$d$_{xy}$  &             ${-0.43}^{1,2}$                             &             \makecell{${0.40}^{1}$\\${-0.28}^{5}$\\${0.26}^{2}$\\${0.25}^{4}$}                       &       ${-0.32}^{2}$                     &             \makecell{${0.40}^{6}$\\${0.28}^{1}$\\${0.26}^{3}$\\${-0.25}^{2}$}             &            \makecell{${0.31}^{3}$\\}              &         \makecell{${0.76}^{1}$\\${0.67}^{2}$\\}             &         \makecell{${-0.42}^{1}$\\${0.37}^{6}$\\${0.27}^{5}$\\}               &          \makecell{${0.61}^{1}$\\${-0.54}^{2}$\\${-0.49}^{6}$\\${-0.33}^{3}$}              \\
 \hline
 Hf  $5$d$_{yz}$  &             \makecell{${0.40}^{2}$\\${0.28}^{4}$\\${0.26}^{1}$\\${-0.25}^{5}$}                             &       ${-0.43}^{4,5}$                             &         \makecell{${-0.30}^{4}$\\${0.30}^{5}$\\${-0.29}^{1}$}                    &      \makecell{${0.40}^{4}$\\${-0.28}^{6}$\\${0.26}^{5}$\\${0.25}^{3}$}                    &          -                &        \makecell{${0.61}^{4}$\\${-0.54}^{5}$\\${-0.49}^{2}$\\${-0.33}^{1}$}              &    \makecell{${0.76}^{4}$\\${0.67}^{5}$\\}                    &        \makecell{${-0.42}^{5}$\\${0.37}^{2}$\\${0.27}^{6}$\\}               \\
 \hline
 Hf  $5$d$_{xz}$  &             \makecell{${0.40}^{3}$\\${-0.28}^{2}$\\${0.26}^{6}$\\${0.25}^{1}$}                            &            \makecell{${0.40}^{5}$\\${0.28}^{3}$\\${0.26}^{4}$\\${-0.25}^{6}$}                       &         -                    &        ${-0.43}^{3,6}$                &      \makecell{${0.36}^{6}$\\${-0.29}^{3}$}                    &       \makecell{${-0.42}^{3}$\\${0.37}^{5}$\\${0.27}^{2}$\\}               &       \makecell{${0.61}^{3}$\\${-0.54}^{6}$\\${-0.49}^{5}$\\${-0.33}^{4}$}                 &            \makecell{${0.76}^{3}$\\ ${0.67}^{6}$\\}           \\
 \hline
  Hf  $5$d$_{x^{2}-y^{2}}$  &       \makecell{${-0.31}^{6}$\\}                                   &          -                          &            -                 &        \makecell{${-0.36}^{3}$\\${0.29}^{6}$}                &            -              &   \makecell{${0.46}^{2}$\\${-0.42}^{1}$\\}                   &        -                &           \makecell{${0.53}^{3}$\\${-0.39}^{6}$\\}           \\
  \hline
   Hf $5$d$_{z^{2}}$  &               \makecell{${0.32}^{1}$\\}                           &             \makecell{${0.30}^{5}$\\${-0.30}^{4}$\\${0.29}^{2}$}                       &           -                  &     -                   &             -             &       \makecell{${-0.37}^{1}$\\}              &        \makecell{${0.54}^{4}$\\${-0.49}^{5}$\\}               &        \makecell{${0.30}^{6}$\\}          \\
   \hline
 Hf  $6$p$_{y}$   &        \makecell{${-0.76}^{2}$\\${-0.67}^{1}$\\}                &             \makecell{${0.61}^{5}$\\${-0.54}^{4}$\\${0.49}^{1}$\\${0.33}^{2}$}             &     \makecell{${-0.37}^{2}$\\}                        &      \makecell{${-0.42}^{6}$\\${0.37}^{4}$\\${0.27}^{1}$\\}                   &            \makecell{${0.46}^{1}$\\${-0.42}^{2}$\\}              &     \makecell{${0.74}^{1,2}$\\${0.56}^{4,5}$}                 &         \makecell{${0.49}^{4}$\\${-0.46}^{1}$\\${-0.45}^{5}$\\${0.27}^{2}$\\${-0.26}^{6}$}               &     \makecell{${-0.49}^{2}$\\${0.46}^{6}$\\${0.45}^{1}$\\${-0.27}^{6}$\\${-0.26}^{4}$}                   \\
 \hline
 Hf  $6$p$_{z}$   &                 \makecell{${-0.42}^{2}$\\${0.37}^{3}$\\${0.27}^{4}$\\}         &        \makecell{${-0.76}^{5}$\\${-0.67}^{4}$\\}                  &        \makecell{${0.54}^{5}$\\${-0.49}^{4}$}                     &         \makecell{${0.61}^{6}$\\${-0.54}^{3}$\\${0.49}^{4}$\\${0.33}^{5}$}                 &        -                  &       \makecell{${-0.49}^{5}$\\${0.46}^{2}$\\${0.45}^{4}$\\${-0.27}^{1}$\\${-0.26}^{3}$}               &          \makecell{${0.74}^{4,5}$\\${0.56}^{3,6}$}              &      \makecell{${0.49}^{3}$\\${-0.46}^{4}$\\${-0.45}^{6}$\\${0.27}^{5}$\\${-0.26}^{1}$}                  \\
 \hline
 Hf  $6$p$_{x}$   &            \makecell{${0.61}^{2}$\\${-0.54}^{1}$\\ ${0.49}^{3}$\\${0.33}^{6}$}              &     \makecell{${-0.42}^{5}$\\${0.37}^{1}$\\${0.27}^{3}$\\}                      &       \makecell{${0.30}^{3}$\\}                      &      \makecell{${-0.76}^{6}$\\${-0.67}^{3}$\\}                  &        \makecell{${0.53}^{6}$\\${-0.39}^{3}$\\}                  &       \makecell{${0.49}^{1}$\\${-0.46}^{3}$\\${-0.45}^{2}$\\${0.27}^{6}$\\${-0.26}^{5}$}               &     \makecell{${-0.49}^{6}$\\${0.46}^{5}$\\${0.45}^{3}$\\${-0.27}^{4}$\\${-0.26}^{1}$}                   &     \makecell{${0.74}^{3,6}$\\${0.56}^{1,2}$}               \\
\end{tabular}
\end{ruledtabular}
\caption{NNN hopping parameters in eV for LDA calculations without SOC. The hopping parameters smaller than $0.25$ eV are not presented in the table.}
\label{tableofnextnearestneighbourhoppingparameters}
\end{table*}

\begin{table*}[h!]
\begin{ruledtabular}
\begin{tabular}{|c|c|c|c|c|c|c|c|c|}\toprule
    & Hf $\tilde{5d_{xy}} up$& Hf $\tilde{5d_{yz}} up$ & Hf $\tilde{5d_{z^{2}}} up$  & Hf $\tilde{5d_{xz}} up$  & Hf $\tilde{5d_{x^{2}-y^{2}}} up$  &Hf $\tilde{6p_{y}} up$  &Hf $\tilde{6p_{z}} up$ &Hf $\tilde{6p_{x}} up$ \\
    \hline
 Sn  $\tilde{5p_{y}} up$   &        \makecell{${-1.17}^{g}$\\$({-1.11}^{d}+$\\$i(-0.013))$\\${-0.49}^{c}$\\${0.47}^{f}$\\${0.42}^{a}+$\\$i(0.011)$}               &         \makecell{${0.48}^{a}$\\${0.26}^{c}$}                  &      \makecell{${0.88}^{d}$\\${-0.86}^{g}$\\${-0.79}^{e}$\\${-0.76}^{b}$}                  &       ${0.81}^{a}$                   &       \makecell{$({-0.96}^{g}+$\\$i(0.015))$\\${0.92}^{d}$}                  &     \makecell{${-1.65}^{g}$\\${-1.42}^{d}$\\$({-0.54}^{a}+$\\$i(-0.034)$)\\${0.38}^{c}$}              &         \makecell{$({-1.04}^{e}+$\\$i(0.015))$\\$({-0.70}^{a}+$\\$i(-0.016))$\\${0.69}^{b}$\\${-0.39}^{c}$}               &           \makecell{$({-0.79}^{d}+$\\$i(-0.016))$\\$({-0.64}^{a}+$\\$i(-0.015))$\\${0.59}^{g}$\\(${-0.38}^{b}+$\\$i(-0.022))$}             \\
 \hline
 Sn  $\tilde{5p_{z}} up$   &          ${0.81}^{a}$              &      \makecell{${-1.17}^{b}$\\${-1.11}^{e}$\\${-0.49}^{g}$\\${0.47}^{d}$\\${0.42}^{a}$}                     &       \makecell{${1.26}^{b}$\\${-1.23}^{e}$\\ ${0.49}^{c}$\\${0.40}^{f}$}                 &         \makecell{${0.48}^{a}$\\${0.26}^{d}$}                &          \makecell{${-0.68}^{f}$\\${-0.59}^{c}$\\${0.30}^{e}$\\${-0.27}^{b}$}              &         \makecell{${-0.79}^{e}$\\${-0.64}^{a}$\\${0.59}^{b}$\\$({-0.38}^{c}+$\\$i(0.020))$}             &            \makecell{${-1.65}^{b}$\\${-1.42}^{e}$\\${-0.54}^{a}$\\${0.38}^{g}$}            &           \makecell{$({-1.04}^{f}+$\\$i(-0.011))$\\${-0.70}^{a}$\\$({0.69}^{c}+$\\$i(0.017))$\\$({-0.39}^{g}+$\\$i(0.012))$}            \\
 \hline
 Sn  $\tilde{5p_{x}} up$   &           \makecell{$({0.48}^{a}+$\\$i(-0.026))$\\${0.26}^{e}$}             &         ${0.81}^{a}$                  &          \makecell{${-0.40}^{c}$\\${0.39}^{d}$\\${0.36}^{f}$\\${0.27}^{g}$}                  &        \makecell{$({-1.17}^{c}+$\\$i(0.013))$ \\${-1.11}^{f}$\\${-0.49}^{b}$\\${0.47}^{e}$\\${0.42}^{a}$}               &       \makecell{${1.22}^{c}$\\${-1.22}^{f}$\\ ${0.72}^{g}$\\$({0.68}^{d}+$\\$i(-0.018))$}                &        \makecell{$({-1.04}^{d}+$\\$i(0.041))$\\$({-0.70}^{a}+$\\$i(0.031))$\\$({0.69}^{g}+$\\$i(0.043))$\\$({-0.39}^{b}+$\\$i(0.025))$}              &         \makecell{${-0.79}^{f}$\\${-0.64}^{a}$\\$({0.59}^{c}+$\\$i(-0.015))$\\$({-0.38}^{g}+$\\$i(0.017))$}              &     \makecell{${-1.65}^{c}$\\${-1.42}^{f}$\\$({-0.54}^{a}+$\\$i(0.027))$\\${0.38}^{b}$}                   \\
\end{tabular}
\end{ruledtabular}
\caption{NN hopping parameters in eV for LDA calculations with SOC. The hopping parameters smaller than $0.25$ eV are not presented in the table.}
\label{tableofnearestneighbourhoppingparameterssoc}
\end{table*}

\begin{table*}[h!]
\begin{ruledtabular}
\begin{tabular}{|c|c|c|c|c|c|c|c|c|}\toprule
    & Hf $\tilde{5d_{xy}} up$ & Hf $\tilde{5d_{yz}} up$ & Hf $\tilde{5d_{z^{2}}} up$& Hf $\tilde{5d_{xz}} up$& Hf $\tilde{5d_{x^{2}-y^{2}}} up$&Hf $\tilde{6p_{y}} up$ &Hf $\tilde{6p_{z}} up$  &Hf $\tilde{6p_{x}} up$\\
    \hline
Hf $\tilde{5d_{xy}} up$  &             ${-0.43}^{1,2}$                              &         \makecell{${0.40}^{1}$\\${-0.28}^{5}$\\${0.26}^{2}$\\${0.25}^{4}$}                           &            \makecell{${-0.32}^{2}$\\}                 &              \makecell{${0.40}^{6}$\\${0.28}^{1}$\\${0.26}^{3}$\\${-0.25}^{2}$}              &          \makecell{${0.31}^{3}$\\}                &        \makecell{${0.75}^{1}$\\${0.67}^{2}$\\}              &           \makecell{${-0.43}^{1}$\\${0.36}^{6}$\\${0.26}^{5}$\\${0.25}^{2}$\\}             &          \makecell{${0.61}^{1}$ \\${-0.54}^{2}$\\${-0.49}^{6}$\\${-0.33}^{3}$\\}             \\
 \hline
Hf $\tilde{5d_{yz}} up$  &             \makecell{${0.40}^{2}$\\${0.28}^{4}$\\${0.26}^{1}$\\${-0.25}^{5}$\\}                             &       ${-0.43}^{4,5}$                            &      \makecell{${-0.31}^{4}$\\${0.30}^{5}$\\${-0.29}^{1}$}                       &      \makecell{${0.40}^{4}$\\${-0.28}^{6}$\\${0.26}^{5}$\\${0.25}^{3}$\\}                    &        -                  &       \makecell{${0.61}^{4}$\\ ${-0.54}^{5}$\\${-0.49}^{2}$\\${-0.33}^{1}$}               &       \makecell{${0.75}^{4}$ \\ ${0.67}^{5}$\\}                 &       \makecell{${-0.43}^{4}$\\${0.36}^{2}$\\${0.26}^{6}$\\${0.25}^{5}$\\}               \\
 \hline
 Hf $\tilde{5d_{xz}} up$  &              \makecell{${0.40}^{3}$\\${-0.28}^{2}$\\${0.26}^{6}$\\${0.25}^{1}$}                           &             \makecell{${0.40}^{5}$\\${0.28}^{3}$\\${0.26}^{4}$\\${-0.25}^{6}$}                      &      -                      &        ${-0.43}^{3,6}$                &          \makecell{${0.36}^{6}$\\$({-0.29}^{3}+$\\$i(-0.011))$}                &     \makecell{\\ ${-0.43}^{3}$\\${0.36}^{5}$\\${0.26}^{2}$\\${0.25}^{6}$\\}                &        \makecell{${0.61}^{3}$\\ ${-0.54}^{6}$\\${-0.49}^{5}$\\${-0.33}^{4}$}                   &           \makecell{${0.75}^{3}$\\${0.67}^{6}$\\}        \\
 \hline
 Hf $\tilde{5d_{x^{2}-y^{2}}} up$  &      \makecell{${-0.31}^{6}$\\}                                    &            -                        &              -               &       \makecell{${-0.36}^{3}$\\$({0.29}^{6}+$\\$i(0.011))$}               &           -               &         \makecell{${0.47}^{2}$\\${-0.41}^{1}$\\}             &          -              &           \makecell{${0.52}^{3}$\\${-0.39}^{6}$\\}          \\
  \hline
  Hf $\tilde{5d_{z^{2}}} up$  &               \makecell{${0.32}^{1}$\\}                           &             \makecell{${0.31}^{5}$\\${-0.30}^{4}$\\${0.29}^{2}$}                       &           -                  &     -                  &             -             &       \makecell{${-0.37}^{1}$\\}              &        \makecell{${0.54}^{4}$\\${-0.50}^{5}$\\}               &        \makecell{${0.31}^{6}$\\}          \\
  \hline
 Hf $\tilde{6p_{y}} up$   &        \makecell{${-0.75}^{2}$\\${-0.67}^{1}$\\}                &             \makecell{${0.61}^{5}$\\${-0.54}^{4}$\\${0.49}^{1}$\\${0.33}^{2}$}             &        \makecell{${-0.37}^{2}$\\}                    &     \makecell{${-0.43}^{6}$\\ ${0.36}^{4}$\\${0.26}^{1}$\\${0.25}^{3}$\\}                 &            \makecell{${0.47}^{1}$\\${-0.41}^{2}$\\}              &     \makecell{${0.70}^{1,2}$\\$({0.52}^{4,5}+$\\$i(0.016))$}                 &       \makecell{${0.49}^{4}$\\$({-0.45}^{5}+$\\$i(-0.018))$\\${-0.43}^{1}$\\$({0.27}^{2}+$\\$i(-0.010))$\\${-0.25}^{6}$}                 &     \makecell{$({-0.49}^{2}+$\\$i(-0.015))$\\${0.45}^{1}$\\$({0.43}^{6}$\\$i(0.013))$\\${-0.27}^{3}$\\${-0.25}^{4}$}                   \\
 \hline
Hf $\tilde{6p_{z}} up$   &                \makecell{${-0.43}^{2}$\\${0.36}^{3}$\\${0.26}^{4}$\\${0.25}^{1}$\\}         &         \makecell{${-0.75}^{5}$\\${-0.67}^{4}$\\}                 &       \makecell{${0.54}^{5}$\\${-0.50}^{4}$\\}                     &         \makecell{${0.61}^{6}$\\${-0.54}^{3}$\\${0.49}^{4}$\\${0.33}^{5}$}                &                          &       \makecell{${-0.49}^{5}$\\$({0.45}^{4}$\\$i(-0.018))$\\${0.43}^{2}$\\$({-0.27}^{1}$\\$i(0.010))$\\${-0.25}^{3}$}               &         \makecell{${0.70}^{4,5}$\\${0.52}^{3,6}$}              &      \makecell{${0.49}^{3}$ \\ ${-0.45}^{6}$\\${-0.43}^{4}$\\${0.27}^{5}$\\$({-0.25}^{2}+$\\$i(-0.016))$}                  \\
 \hline
 Hf $\tilde{6p_{x}} up$   &            \makecell{${0.61}^{2}$\\${-0.54}^{1}$\\ ${0.49}^{3}$ \\${0.33}^{6}$}            &     \makecell{${-0.43}^{5}$\\${0.36}^{1}$\\${0.26}^{3}$\\${0.25}^{4}$\\}                    &         \makecell{${0.31}^{3}$\\}                    &       \makecell{${-0.67}^{3}$\\${-0.75}^{6}$\\}                  &        \makecell{${0.52}^{6}$\\${-0.39}^{3}$\\}                  &       \makecell{$({0.49}^{1}+$\\$i(0.016))$\\${-0.45}^{2}$\\$({-0.43}^{3}+$\\$i(-0.013))$\\${0.27}^{6}$\\${-0.25}^{5}$}               &        \makecell{${-0.49}^{6}$\\${0.45}^{3}$\\${0.43}^{5}$\\${-0.27}^{4}$\\$({-0.25}^{1}+$\\$i(-0.016)$)}             &      \makecell{$({0.70}^{3,6}+$\\$i(0.014))$\\${0.52}^{1,2}$}               \\
 
\end{tabular}
\end{ruledtabular}
\caption{NNN hopping parameters in eV for LDA calculations with SOC. The hopping parameters smaller than $0.25$ eV are not presented in the table.}
\label{tableofnextnearestneighbourhoppingparameterssoc}
\end{table*}

\end{document}